\documentclass[a4paper,fleqn,usenatbib,twocolumn]{mnras}

\usepackage{newtxtext,newtxmath}
\usepackage[T1]{fontenc}
\usepackage{ae,aecompl}

\usepackage{graphicx}	% Including figure files
\usepackage{amsmath}	% Advanced maths commands
\usepackage{amssymb}	% Extra maths symbols
\usepackage[usenames,dvipsnames]{color} % Allows for comments

\def\code#1{\texttt{#1}}

\definecolor{purple}{rgb}{0.5,0,0.5}
\definecolor{darkgreen}{rgb}{0.1,0.6,0.1}
\definecolor{orange}{rgb}{1,0.6,0}

\title[Differential Rotation in Eclipsing Binaries]{Differential Rotation in Convective Envelopes: Constraints from Eclipsing Binaries}

\author[Adam S. Jermyn, Jamie Tayar and Jim Fuller]{
Adam S. Jermyn$^{1,2}$\thanks{E-mail: adamjermyn@gmail.com},
 Jamie Tayar$^{3,4}$ and Jim Fuller$^{5}$\\
$^{1}$Center for Computational Astrophysics, Flatiron Institute, New York, NY 10010, USA\\
$^{2}$Kavli Institute for Theoretical Physics, University of California at Santa Barbara, Santa Barbara, CA 93106, USA\\
$^{3}$Institute for Astronomy, University of Hawaii, 2680 Woodlawn Drive, Honolulu, Hawaii 96822, USA\\
$^{4}$Hubble Fellow\\
$^{5}$TAPIR, Mailcode 350-17, California Institute of Technology, Pasadena, CA 91125, USA
}

\date{Accepted XXX. Received YYY; in original form ZZZ}

\pubyear{2019}

\begin{document}
\label{firstpage}
\pagerange{\pageref{firstpage}--\pageref{lastpage}}
\maketitle

\begin{abstract}
Over time, tides synchronize the rotation periods of stars in a binary system to the orbital period.
However, if the star exhibits differential rotation then only a portion of it can rotate at the orbital period, so the rotation period at the surface may not match the orbital period.
The difference between the rotation and orbital periods can therefore be used to infer the extent of the differential rotation.
We use a simple parameterization of differential rotation in stars with convective envelopes in circular orbits to predict the difference between the surface rotation period and the orbital period.
Comparing this parameterization to observed eclipsing binary systems, we find that in the surface convection zones of stars in short-period binaries there is very little radial differential rotation, with $|r\partial_r \ln \Omega| < 0.02$.
This holds even for longer orbital periods, though it is harder to say which systems are synchronized at long periods, and larger differential rotation is degenerate with asynchronous rotation.

\end{abstract}


%%%%%%%%%%%%%%%%%%%%%%%%%%%%%%%%%%%%%%%%%%%%%%%%%%

%%%%%%%%%%%%%%%%% BODY OF PAPER %%%%%%%%%%%%%%%%%%

\section{Introduction}

Despite much work there remains significant uncertainty as to the primary location of rotational shear in stars.
In particular, whether shear is strongest in radiative or convective regions remains an open question~\citep{2014ApJ...788...93C,2015ApJ...808...35K,2019A&A...626L...1E}, and one which has significant consequences for the spin periods of compact objects~\citep{2017ApJS..232...23H}.
In the Sun, helioseismic inversions provide a measurement of the shear in the convection zone and place some constraints on the upper parts of the radiative interior, though the rotation of the deep interior remains uncertain~\citep{1998ApJ...505..390S,2010ApJ...720..494A}.

Eclipsing binary systems provide a precision laboratory for probing stellar structure.
Their masses, radii and orbital parameters may be precisely determined from the timing and depths of the eclipses~\citep{2004MNRAS.351.1277S,2010A&ARv..18...67T}.
When combined with models these data may be used to constrain fundamental parameters of stellar structure and evolution~\citep{2017ApJ...849...18C}.

Such constraints are particularly useful in the domain of angular momentum transport, and a number of efforts have been made to study rotation in eclipsing binary systems.
\citet{2010A&ARv..18...67T} compared $v\sin i$ with its pseudosynchronization value to find that many close binary systems are indeed (pseudo)-synchronized.
\citet{2014ApJ...785....5G} combined spot timing with orbital measurements of {\emph Kepler} eclipsing binaries to constrain synchronization of red giants.
More recently,~\citet{2017AJ....154..250L} used the same kinds of data to infer latitudinal differential rotation in solar-type stars.

\begin{figure}
\includegraphics[width=0.47\textwidth]{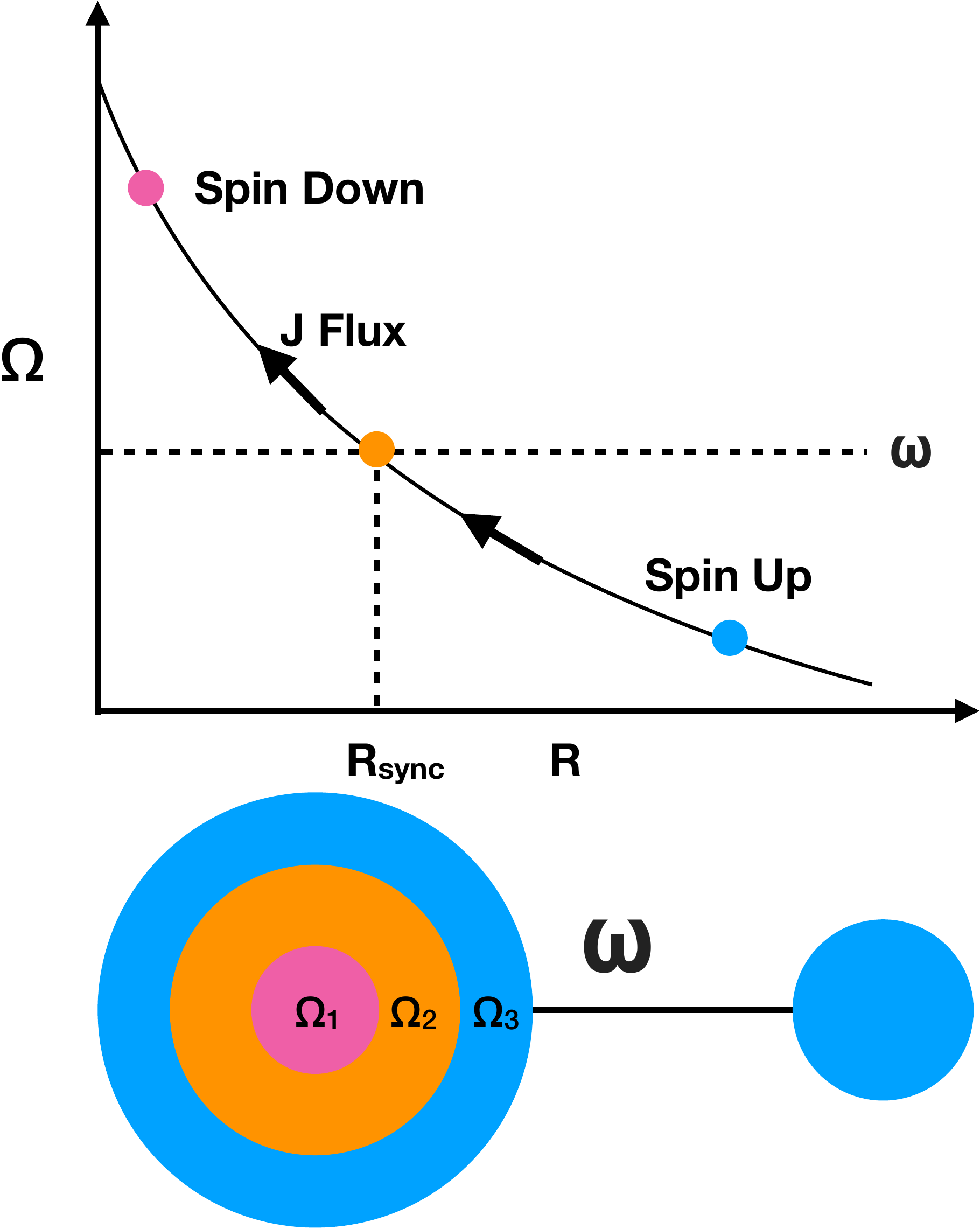}
\caption{Upper: Different regions of a differentially rotating star rotate at different rates, so not all parts of the star can rotate synchronously with the orbital rate $\omega$. In equilibrium angular momentum must flow from regions that are super-synchronous to those which are sub-synchronous. For simplicity only the radial variation of the rotation rate $\Omega$ is shown. Lower: A side view of the binary is shown.}
\label{fig:schema0}
\end{figure}

If a star is differentially rotating, then not all of the star can be rotating synchronously with its orbit.
This is shown schematically in Figure~\ref{fig:schema0}.
Some regions must be spinning sub-synchronously, and tides act to spin these up.
Likewise, other regions must be spinning super-synchronously; tides act to slow these down.
In equilibrium the net torque on the star vanishes, and so the star as a whole can be said to be rotating synchronously with the orbit.
This means that in equilibrium the surface of the star is likely {\emph not} synchronized with the orbit.
Rather, the surface is either sub- or super-synchronous, and it is only deeper down that the star corotates with the orbit.

We compute the tidal torque on one component of a binary using a simplified model of its internal rotation profile.
Setting the net torque to zero allows us to compute the ratio of the surface rotation period to the orbital period as a function of the differential rotation in the star.
Assuming that the differential rotation profile is similar for stars with similar rotation periods, we infer both the radial and latitudinal components of the shear from the sample of~\citet{2017AJ....154..250L}.
The stars in this sample are of the right spectral types to have significant outer convection zones.
Convection redistributes angular momentum quickly, likely on a time-scale of order the convective turnover time~\citep{2013MNRAS.431.2200L}, and so the tides only serve to set the total amount of angular momentum, not its ultimate distribution.
Hence our findings for these systems likely generalize to single stars with comparable rotation periods.

We begin in Sections~\ref{sec:torque} and~\ref{sec:lag} by reviewing the tidal torque on a star and deriving the condition of synchronization.
In Section~\ref{sec:profile} we introduce our parameterized rotation profile.
We describe our data processing and analysis in Sections~\ref{sec:data},~\ref{sec:models}, and~\ref{sec:infer}.
We then explore the results of the analysis in Section~\ref{sec:res} and discuss the implications in Section~\ref{sec:disc}.

To briefly summarize our results, we find that at short periods the radial relative shear must be small, and in particular that it is too small to account for the seismically-inferred rotation periods of stellar cores~\citep{2015ApJ...808...35K}.
We likewise constrain the latitudinal shear to be small, though the uncertainties in this are larger than for the radial case.

\section{Tidal Torque}
\label{sec:torque}

\begin{figure}
\includegraphics[width=0.47\textwidth]{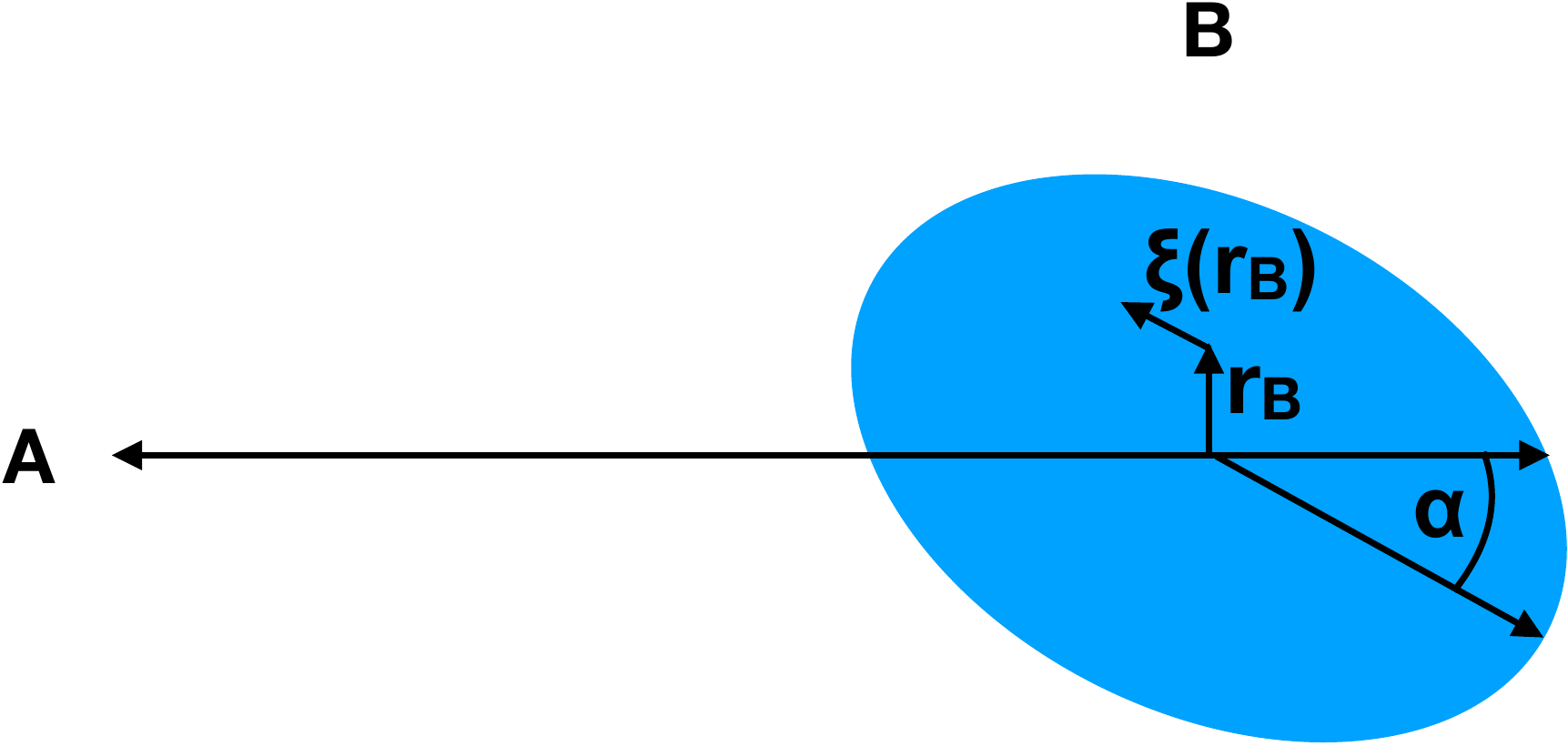}
\caption{Two stars labeled $A$ and $B$ are shown from above in a circular orbit. $r_A$ and $r_B$ are the spherical radial coordinates referenced to the centers of stars $A$ and $B$ respectively. Star $B$ forms a tidal bulge in response to the potential of star $A$. The bulge is parameterized by the displacement field $\boldsymbol{\xi}(\boldsymbol{r})$, which specifies the displacement of a fluid element which would be located at $\boldsymbol{r}_B$ in the absence of the tidal potential. The bulge lags by an angle $\alpha$ relative to the line joining the two stars.}
\label{fig:schema2}
\end{figure}

Consider a binary system with stars labeled $A$ and $B$ in a circular orbit about one another, as shown in Figure~\ref{fig:schema2}.
Fluid elements in star $B$ have position $\boldsymbol{r}_B$ relative to the center of that star.
In the frame co-rotating with the binary orbit the position of a fluid element rotates, such that
\begin{align}
	\frac{d\boldsymbol{r}_{B}}{dt} &= (\Omega-\omega) \hat{z}\times \boldsymbol{r}_{B},
	\label{eq:r_rot}
\end{align}
where $\hat{z}$ is the unit vector along the rotation axis of the star, which we assume to be aligned with the orbital axis, $\Omega$ is the local angular velocity of the fluid relative to the center of its star and $\omega$ is the orbital angular velocity.
Note that we assume $\Omega$ to be time-independent for each Lagrangian fluid element, so that $\boldsymbol{r}_{B}(t)$ describes an element undergoing rotation at fixed angular velocity.

\begin{figure}
\centering
\includegraphics[width=0.4\textwidth]{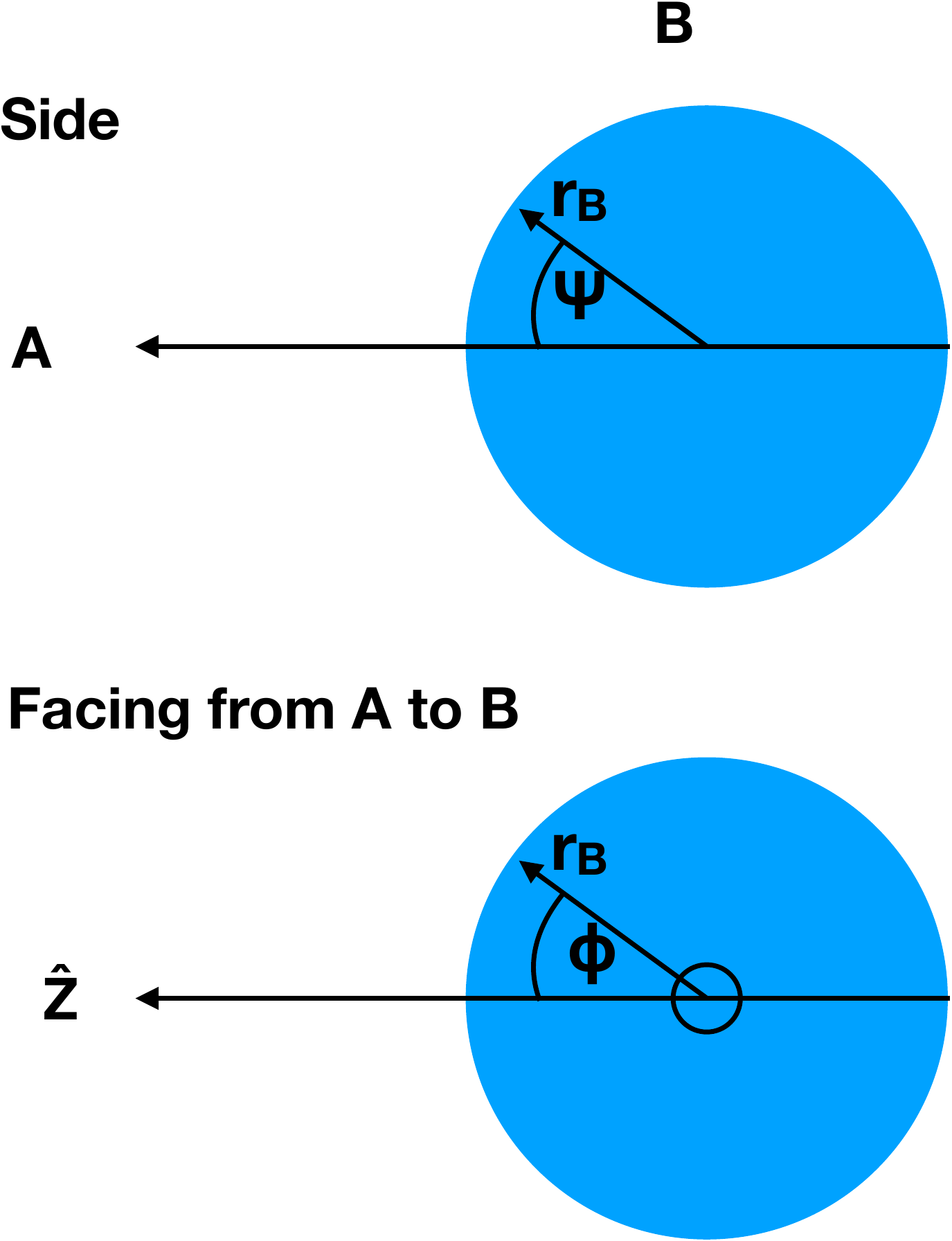}
\caption{The angle $\psi$ is that between the vector $-\hat{r}_{A\rightarrow B} = \hat{r}_{B\rightarrow A}$ and $\boldsymbol{r}_B$. The angle $\phi$ is that between $\boldsymbol{r}_B$ and the orbital axis $\hat{z}$. The small black circle in the lower schematic indicates that in that view $\hat{r}_{B\rightarrow A}$ runs out of the page.}
\label{fig:schema5}
\end{figure}

The gravitational pull of one of the stars may be expanded about the center of the other to find the tidal potential.
To leading order this is a quadrupole, so in the frame co-rotating with the binary orbit
\begin{align}
	\delta \Phi_{A\rightarrow B} = \frac{G M_A r_B^2}{a^3} Y_{2,0}(\psi,\phi),
	\label{eq:potential}
\end{align}
where $\Phi_{A\rightarrow B}$ is the tidal potential felt by star $B$ owing to star $A$, $M_A$ is the mass of star~$A$, $a$ is the binary separation, $Y_{lm}$ are the spherical harmonics and the angles $\psi$ and $\phi$ are the usual spherical coordinates, referenced to the tidal axis, and are shown in Figure~\ref{fig:schema5}.

In response to the tidal potential star $B$ will form a tidal bulge.
The full calculation of the tidal displacement is complicated, and we refer the reader to the more complete treatments of~\citet{1975A&A....41..329Z},~\citet{1977A&A....57..383Z} and~\citet{2014ARA&A..52..171O}.
For our purposes it suffices to parameterize the bulge by the displacement field $\boldsymbol{\xi}(\boldsymbol{r}_B)$, which specifies the displacement of a fluid element which would be located at $\boldsymbol{r}_B$ in the absence of the tidal potential, as shown in Figure~\ref{fig:schema2}.
In hydrostatic equilibrium this bulge would be aligned with the axis between the two stars.
We denote this equilibrium tide by $\boldsymbol{\xi}_{\rm eq}$, and note that it obeys
\begin{align}
	\label{eq:xi_eq}
	\xi_{{\rm eq}, r} &= -\frac{\delta \Phi_{A\rightarrow B}}{g}\\
	\intertext{and}
	\label{eq:xi_eq2}
	\nabla\cdot\boldsymbol{\xi}_{\rm eq} &= 0
\end{align}
\citep{2012A&A...544A.132R}, where $g$ is the acceleration due to the gravity of star~$B$.
Note that because $g$ and $\delta \Phi_{A\rightarrow B}$ both vary on the length-scale $r_B$, the non-radial component of the displacement ($\xi_{{\rm eq}, \perp}$) is of order $\xi_{{\rm eq}, r}$.

The specific torque on each fluid element is
\begin{align}
	\boldsymbol{\tau} = \boldsymbol{\xi} \times \nabla \delta \Phi_{A\rightarrow B}.
\end{align}
By symmetry the net torque is along $\hat{z}$, so we may write the local contribution to the net torque as
\begin{align}
	\boldsymbol{\tau} \approx \hat{z} \xi_{\rm eq} |\nabla\delta \Phi_{A\rightarrow B}| \approx \xi_{{\rm eq}, r} \frac{\delta \Phi_{A\rightarrow B}}{r_B} \approx \frac{(\delta \Phi_{A\rightarrow B})^2}{g r_B} \approx \left(\frac{G M_A}{a^3}\right)^2 \frac{r_B^3}{g},
	\label{eq:tau0}
\end{align}
where we have used the fact that $\delta \Phi_{A\rightarrow B}$ varies on the length-scale $r_B$.

When the tidal bulge is aligned with the axis connecting the two stars the torque in equation~\eqref{eq:tau0} vanishes when integrated over star~$B$.
Turbulent and dissipative processes may force the bulge to lag or lead relative to that line.
To parameterize this we take the tidal displacement to obey
\begin{align}
	\boldsymbol{\xi}(\boldsymbol{r}_B, t) = \boldsymbol{\xi}_{\rm eq}\left[\boldsymbol{r}_{B}\left(t - \frac{\alpha}{\Omega-\omega}\right)\right],
	\label{eq:xi}
\end{align}
where we have introduced the lag angle $\alpha$ shown in Figure~\ref{fig:schema2}.
This angle is just the phase difference between the local rotation of the fluid and the displacement.
So, in general, it may be a function of $\boldsymbol{r}_B$, as different parts of the star may rotate at different rates and experience different degrees of dissipation.

When the angle is non-zero there is a torque proportional to $\sin\alpha$, which we approximate in the limit of small lag angle by $\alpha$.
So the contribution of each fluid element to the net torque is of order
\begin{align}
	\boldsymbol{\tau} \approx \hat{z} \alpha \left(\frac{G M_A}{a^3}\right)^2 \frac{r_B^3}{g}.
\end{align}
In the case where $\alpha$ is only a function of radius and not of latitude or longitude on the star we may integrate this over the whole star to find the net torque
\begin{align}
	\mathcal{T} = \int \boldsymbol{\tau} dm \approx 4\pi \hat{z} \left(\frac{G M_A}{a^3}\right)^2 \int_0^{R_B}  \rho \alpha \frac{r_B^5}{g}  dr_B,
\end{align}
where $\rho$ is the unperturbed density of the material in star~$B$ and $R_B$ is the outer radius of star $B$ in the absence of tidal perturbations.

More generally we are interested in the case where $\alpha$ depends on latitude from the rotation axis.
In this case the derivation is more complicated, and we provide the details in Appendix~\ref{appen:torque}.
Briefly, we compute the equilibrium tide explicitly and assume that $\alpha$ and $\Omega$ are uniform along the path that material follows as it rotates.
Some algebra then yields the result given by equation~\eqref{eq:T3}, namely
\begin{align}
	\mathcal{T} &\propto\hat{z}\int_0^{R_B}dr_B \int_0^{\pi}d\psi \sin\psi \int_0^{2\pi} d\phi \rho \alpha \frac{r_B^5}{g}\nonumber\\
&\times\left[\cos\psi\left(\cos\psi - 3\cos^3\psi - 3\cos(2\psi)\sin\psi\right)\right.\nonumber\\
&\left.-\frac{1}{4}\sin^2\phi \sin\psi \big(6\cos\psi + 6\cos(3\psi) + \sin\psi - 3 \sin(3\psi)\big)\right].
\label{eq:torque0}
\end{align}
Because we are interested in the case of synchronous rotation, for which there is no net torque, we have dropped constant factors and all dependence on $a$, because these are not sensitive to the structure and rotation of star~$B$.

\section{Lag Angle and $Q$}
\label{sec:lag}

The lag angle is related to the tidal quality factor $Q$ by~\citep{1966Icar....5..375G}
\begin{align}
	|\alpha| \approx \frac{1}{2Q}.
\end{align}
If the orbital frequency $\omega$ exceeds the stellar spin frequency $\Omega$ then the bulge lags behind the companion star and $\alpha > 0$.
Otherwise, the bulge runs ahead of the companion and $\alpha < 0$.
Hence,
\begin{align}
	\alpha \approx \frac{\mathrm{sign}(\omega-\Omega)}{2Q}.
\end{align}

The quality factor $Q$ is the ratio of the energy in the tidal bulge to the energy dissipated over one tidal period.
The kinetic energy of the bulge is small relative to its potential energy.
The latter vanishes at first order in $\boldsymbol{\xi}$, has characteristic scale set by the potential $\Phi_B$ of star~$B$, and varies over the length-scale $r_B$, so
\begin{align}
	E \approx \xi^2 \frac{\Phi_B}{r_B^2} \approx \xi^2 \frac{g}{r_B},
\end{align}
where $g$ is the local acceleration due to the gravity of star~$B$.
Turbulent viscosity dissipates energy, so that
\begin{align}
	\frac{dE}{dt} \approx \dot{\boldsymbol{\xi}}\cdot\left(\nu_{\rm c} \nabla^2 \dot{\boldsymbol{\xi}}\right) \approx \xi^2 |\omega-\Omega|^2 \frac{\nu_{\rm c}}{r_B^2},
\end{align}
where $\nu_{\rm c}$ is the turbulent convective viscosity~\citep{1977Icar...30..301G} and we have approximated time derivatives as producing factors of the tidal frequency and spatial derivatives as producing factors of $r_B^{-1}$.
With this we find~\citep{1974Icar...23...42H}
\begin{align}
	Q \approx \frac{E |\omega-\Omega|}{dE/dt} \approx \frac{g r_B}{|\omega-\Omega| \nu_{\rm c}}.
\end{align}

The turbulent viscosity may be estimated as
\begin{align}
	\nu_{\rm c} \approx \frac{h u_c}{1 + \left(\frac{\omega-\Omega}{u_c/h}\right)^2},
	\label{eq:nuc}
\end{align}
where $h$ is the pressure scale-height and $u_c$ is the convective velocity.
The systems of interest are nearly synchronous, so $|\omega-\Omega| \ll \omega$.
Moreover, because of the factor of $r_B^5/g$ in equation~\eqref{eq:torque0}, the tidal torque is dominated by the outermost regions of the star.
These are the ones which undergo the fastest convection, and for all of the systems we shall examine these regions have $u_c/h > \omega \gg |\omega-\Omega|$.
Hence, we may simplify equation~\eqref{eq:nuc} to just
\begin{align}
	\nu_{\rm c} \approx h u_c,
\end{align}
and thereby find
\begin{align}
	\alpha \propto (\omega-\Omega) \frac{u_c h}{g r_B}.
\end{align}

Putting it all together and requiring a zero net tidal torque we find
\begin{align}
0 &= \hat{z}\int_0^{R_B}dr_B \int_0^{\pi}d\psi \sin\psi \int_0^{2\pi} d\phi (\omega-\Omega) h u_c \rho \frac{r_B^4}{g^2}\nonumber\\
&\times\left[\cos\psi\left(\cos\psi - 3\cos^3\psi - 3\cos(2\psi)\sin\psi\right)\right.\nonumber\\
&\left.-\frac{1}{4}\sin^2\phi \sin\psi \left(6\cos\psi + 6\cos(3\psi) + \sin\psi - 3 \sin(3\psi)\right)\right],
\end{align}
Because $h$, $\rho$ and $u_c$ are thermodynamic properties of the star they are principally functions of $r$.
Likewise, $g$ is nearly independent of $\psi$ and $\phi$.
So our synchronization criterion becomes
\begin{align}
0 &= \hat{z}\int_0^{R_B}dr_B h u_c \rho \frac{r_B^4}{g^2} \int_0^{\pi}d\psi \sin\psi \int_0^{2\pi} d\phi (\omega-\Omega) \nonumber\\
&\times\left[\cos\psi\left(\cos\psi - 3\cos^3\psi - 3\cos(2\psi)\sin\psi\right)\right.\nonumber\\
&\left.-\frac{1}{4}\sin^2\phi \sin\psi \big(6\cos\psi + 6\cos(3\psi) + \sin\psi - 3 \sin(3\psi)\big)\right],
	\label{eq:torque_balance}
\end{align}

In order for equation~\eqref{eq:torque_balance} to hold it must be that some parts of the star are rotating at super-synchronous rates while others are rotating sub-synchronously.
There must then be angular momentum transported within the star to ensure local angular momentum equilibrium, as shown in Figure~\ref{fig:schema0}.
This is readily achieved: angular momentum is transported by the convection zone on a time-scale of $h/u_c$ which is much faster than the synchronization time, so the star is in angular momentum equilibrium.

For the same reason we do not need to worry about the rotation profile of the star being different from that of a single star: convective angular momentum transport acts so much faster than tidal torques that the profile should be nearly the same as that in the absence of tides.
As such tides only affect the total angular momentum, not its distribution within the convection zone.

\section{Rotation Profile}
\label{sec:profile}

Using equation~\eqref{eq:torque_balance} combined with a prescription for the spatial variation of $\Omega$ we may relate the surface rotation rate to the orbital frequency.
For this we choose a simple profile of the form
\begin{align}
	\Omega(r,\theta) = \Omega_0 \left(\frac{r}{R_\star}\right)^{\beta} \big(1 + c_2 P_2(\cos\theta)\big),
	\label{eq:rot_prof}
\end{align}
where $\Omega_0$ sets the overall scale of the rotation, the first factor controls the radial shear and the second controls the latitudinal component.
Here $\theta$ is the angle from the stellar rotation axis, which is related to $\psi$ and $\phi$ by
\begin{align}
	\cos\theta = \sin\psi\cos\phi.
\end{align}
The radial dependence of this model is inspired by various theoretical arguments suggesting that rotation in convection zones ought to behave as a power-law in radius~\citep{2013MNRAS.431.2200L,2015ApJ...808...35K}\footnote{For simplicity we have treated this as a spherical radial dependence. Changing to a cylindrical dependence would only introduce new latitudinal factors. We have already included the lowest order non-vanishing spherical harmonic contribution, so we neglect these additional factors.}.
Neglecting the tidal perturbation, the angular dependence of $\Omega$ is just the leading order term which is consistent with all of the symmetries of a rotating star\footnote{This neglect is justified if the tidal torque does not significantly change the rotation rate over one orbit, which is generally the case.}.
This form is also consistent with theoretical predictions~\citep{1999A&A...344..911K} as well as helioseismic inversions~\citep{1998ApJ...505..390S,2010ApJ...720..494A}.

\begin{figure}
	\includegraphics[width=0.5\textwidth]{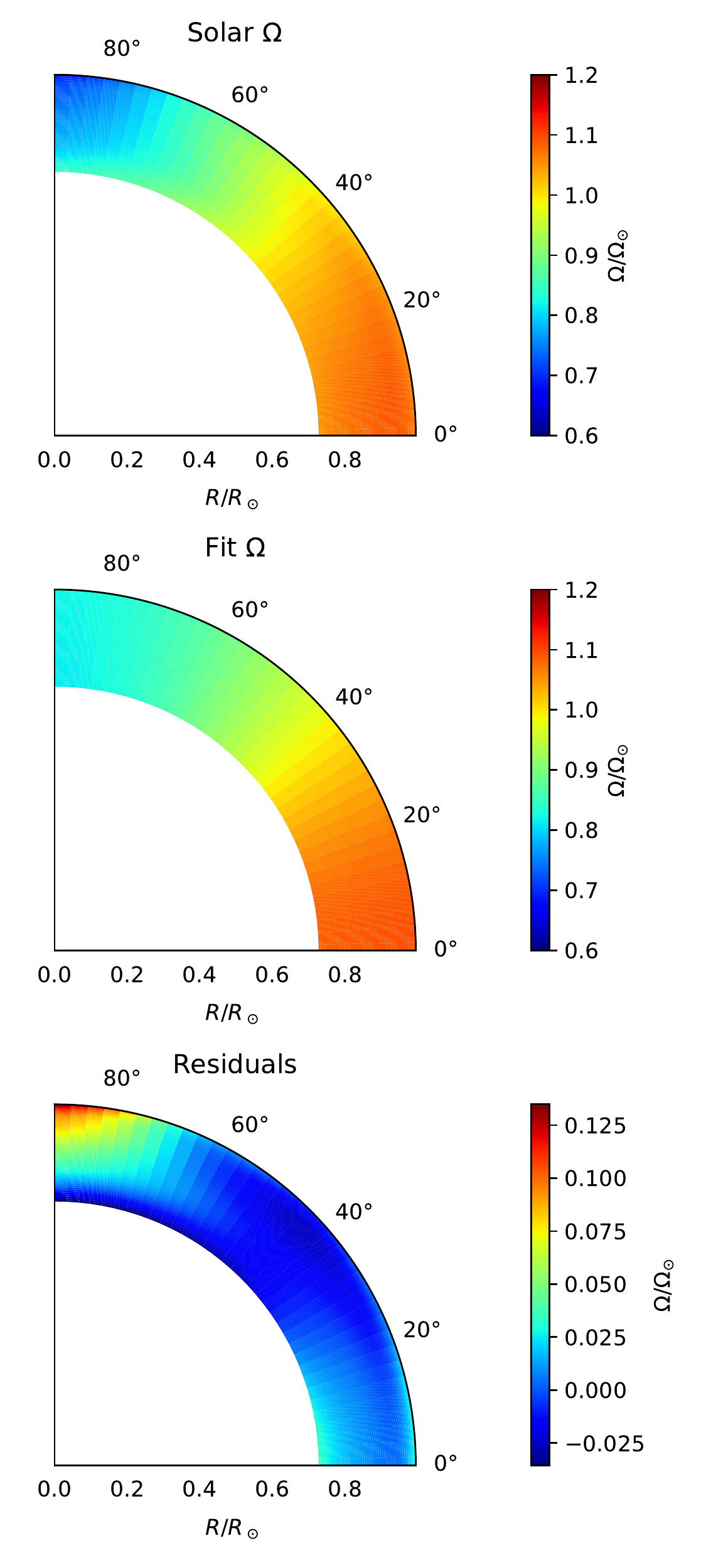}
	\caption{(Top) The rotation profile of the solar convection zone from~\citet{2008ApJ...681..680A}, normalized to the volume-averaged solar rotation rate. (Middle) The rotation profile in equation~\eqref{eq:rot_prof} fitted to the solar profile. (Bottom) The residuals between the data and the fit. The angle labels mark latitude. The best fit parameters are $\beta=0.03$, $c_2=-0.18$ and $\Omega_0 = 1.004 \Omega_\odot$.}
	\label{fig:solar}
\end{figure}

The solar rotation profile provides a natural test of this prescription.
A helioseismic inversion of the rotation profile in the solar convection zone was obtained from~\citet{Private}, corresponding to that appearing in~\citet{2008ApJ...681..680A}.
Equation~\eqref{eq:rot_prof} was fit to the solar profile by minimizing the volume-weighted squared difference in $\Omega$.
The result is shown in Figure~\ref{fig:solar}.
The best fit parameters are $\beta=0.03$, $c_2=-0.18$ and $\Omega_0 = 1.004 \Omega_\odot$, which result in a root-mean-squared error of $1$~per-cent.
Except near the poles the fit is very good, with residuals less than $3$~per-cent.
Toward the poles the fit worsens, but neither the tides nor starspots are expected to be sensitive to that region, and that is the region where the helioseismic inversion is most uncertain.

Inserting equation~\eqref{eq:rot_prof} into equation~\eqref{eq:torque_balance} and performing the integration over $\psi$ and $\phi$ we find
\begin{align}
	\frac{\omega}{\Omega_0} = k_\star(\beta) \left(1 - \frac{2}{7}c_2\right),
	\label{eq:torque_balance2}
\end{align}
where
\begin{align}
	k_\star(\beta) \equiv \frac{\int_0^{R_B}  \frac{r_B^{4+\beta}}{g} \rho h u_c  dr_B}{\int_0^{R_{\star}} R_B^\beta \frac{r_B^{4}}{g} \rho h u_c dr_B}.
\end{align}
is a function only of $\beta$ and the structure of the star.
For simplicity, we treat $g \propto r^{-2}$, having already made this approximation in deriving the tidal torque, and obtain
\begin{align}
	k_\star(\beta) \equiv \frac{\int_0^{R_B}  r_B^{6+\beta} \rho h u_c  dr_B}{\int_0^{R_{\star}} R_B^\beta r_B^{6} \rho h u_c dr_B}.
	\label{eq:kstar}
\end{align}

The surface rotation rates in our sample come from spot measurements.
Assuming a typical spot latitude of $30^\circ$, the surface rotation rate at the spot is
\begin{align}
	\Omega_{\rm s} &= \Omega_0 \left[1 + c_2 P_2\left(\cos\frac{\pi}{6}\right)\right]\\
	&= \Omega_0 \left(1 + \frac{5}{8}c_2 \right).
\end{align}
So
\begin{align}
	\frac{\Omega_{\rm s}}{\omega} = \frac{P_{\rm orb}}{P_{\rm s}} \approx k^{-1}_\star(\beta)\frac{1 + \frac{5}{8}c_2}{1 - \frac{2}{7}c_2}.
	\label{eq:omegas}
\end{align}

\section{Data}
\label{sec:data}

We obtained rotation periods and orbital periods for 816 {\emph{Kepler}} eclipsing binary systems from~\citet{2017AJ....154..250L}.
We excluded triple-star systems from these samples using the catalogs of~\citet{2013ApJ...768...33R} and~\citet{2016MNRAS.455.4136B}.
We further exclude the ``false positive'' systems described by~\citet{2016MNRAS.455.4136B} because, despite not being triple systems, these have unusual light curves which could interfere with the inference of rotation periods.

Using equations (1) and (2) of~\citet{2017AJ....154..250L} as well as eclipse timing from the Kepler Eclipsing Binary catalog~\citep{2012AJ....143..123M,2014PASP..126..914C,2015MNRAS.452.3561L,2016AJ....151..101A} we estimated the eccentricities of this sample and excluded all systems with $e > 0.1$.
This ensures that the synchronous rotation rate in the absence of differential rotation is close to the orbital period for all remaining systems.

\section{Stellar Models}
\label{sec:models}

\begin{figure}
\includegraphics[width=0.5\textwidth]{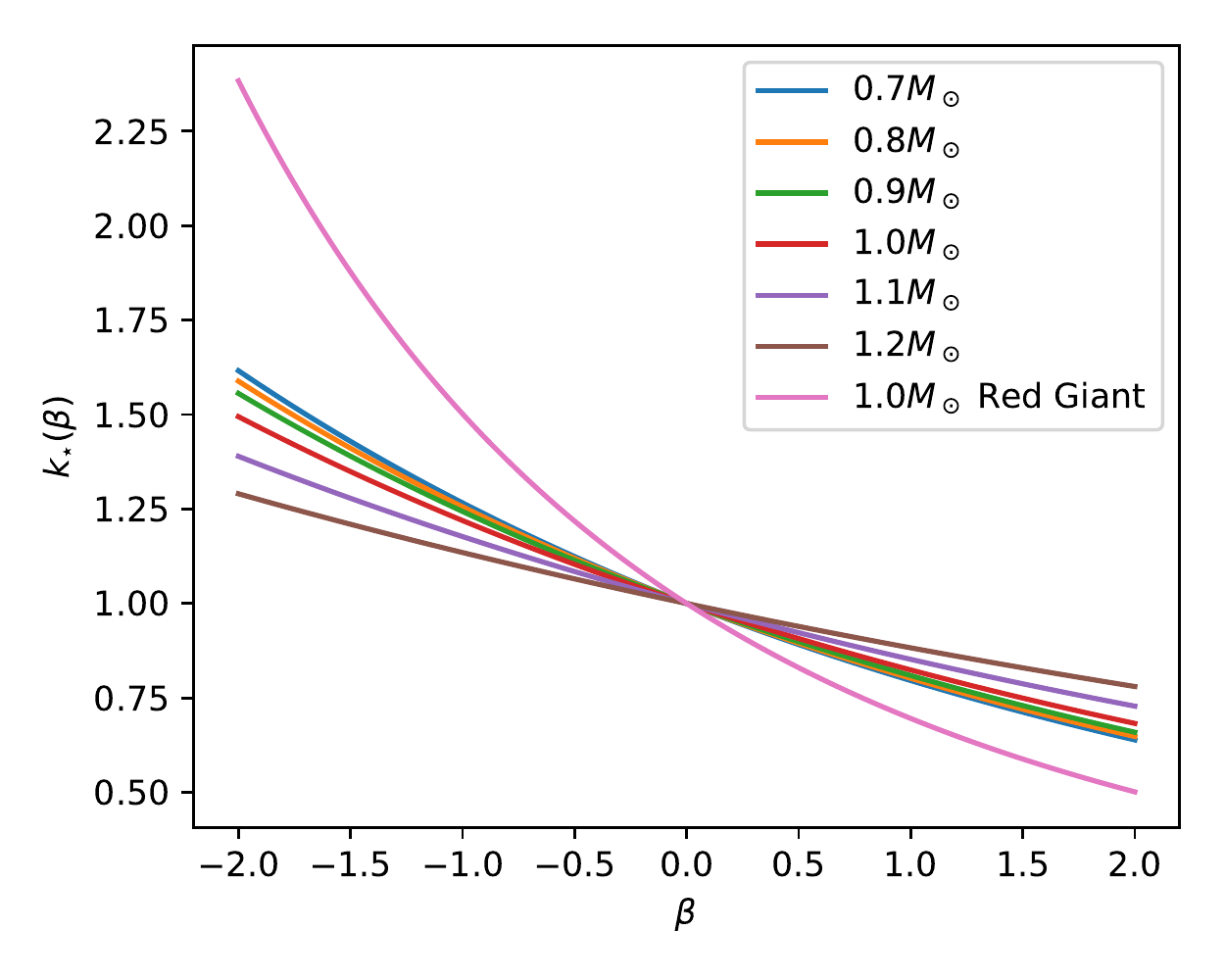}
\caption{The ratio of the surface rotation rate to the orbital rotation rate, $k_*(\beta)$, is shown as a function of radial shear $\beta$ for main-sequence stellar models with masses of $0.7$, $0.8$, $0.9$, $1.0$, $1.1$ and $1.2 M_\odot$.}
\label{fig:kstar}
\end{figure}

Following~\citet{2017AJ....154..250L}, we assume that our sample is mostly comprised of main-sequence stars.
We evaluated $k_\star(\beta)$ for several main-sequence stellar models from $0.7 M_\odot$ to $1.2 M_\odot$.
We treat the lag angle as zero in radiative zones and so only integrate over the convection zones\footnote{This is also justified if there is minimal shear in radiative zones. Some observations point in this direction, e.g.~\citet{10.1093/mnras/stz1171} find minimal differential rotation in the radiative zones of six $\gamma$ Doradus stars.}.
Figure~\ref{fig:kstar} shows $k_\star(\beta)$ computed for main-sequence stellar models with masses of $0.7$, $0.8$, $0.9$, $1.0$, $1.1$ and $1.2 M_\odot$, as well as a $1 M_\odot$ red giant at an age of $12{\rm GYr}$.
The variation in $k_\star$ with $\beta$ becomes larger as the convection zone deepens.
This is because $r$ varies more over a deeper convection zone, so the effect of the power law profile is larger.

Note that $k_\star$ is not sensitive to the overall scale of the stellar radius, mass, or temperature, though it is sensitive to the relative profiles of these.
Hence, for stars with similar structures $k_\star$ ought to be similar.
Moreover, the overall dynamic range is small, so it suffices to use just our representative sample of stellar models.
The stars in our sample are mostly solar-type, so for each star we use $k_\star(\beta)$ computed using the main sequence model which most closely matches the $T_{\rm eff}$ of that star.
We use the median $T_{\rm eff}$ reported in the Kepler Eclipsing Binary catalog for each star.

These calculations used revision 11701 of the Modules for Experiments in Stellar Astrophysics
\citep[\code{MESA}][]{Paxton2011, Paxton2013, Paxton2015, Paxton2018, Paxton2019}.
Details of the microphysical inputs are given in Appendix~\ref{appen:mesa}.

Models were created on the main sequence and evolved from there.
All other parameters were set to their default values.
In particular, no convective overshoot was used, and a simple atmospheric boundary condition approximating optical depth $\tau=2/3$ was used.
The inlists, processing scripts, and model output are available at~\url{Zenodo.org}.

\section{Inference}
\label{sec:infer}

The data provide access to $P_{\rm orb}/P_{\rm s}$.
Our aim is to infer $\beta$ and $c_2$.
Unfortunately, these are degenerate, so we cannot infer them both, or indeed either, without more information.

Fortunately~\citet[][Figure 12]{2017AJ....154..250L} infer $|c_2|$ from the existence of multiple peaks in the periodograms of many stars in their sample.
The data appear to follow a rough power-law trend in period.
That is, we suggest that their data may be modelled by
\begin{align}
	|c_2| \propto\left(\frac{P_{\rm s}}{10{\rm d}}\right)^{\gamma},
	\label{eq:dlnO}
\end{align}
where $\gamma$ is of order $1$.
This dependence on period lifts the degeneracy and makes it possible to infer $\beta$, though at the cost that the data are not as informative about $c_2$.

Inspired by equation~\eqref{eq:dlnO}, we perform our inference assuming that
\begin{align}
	c_2 = \lambda \left(\frac{P_{\rm s}}{10{\rm d}}\right)^{\gamma},
	\label{eq:dlnO2}
\end{align}
where $\lambda$ and $\gamma$ are taken to be universal across systems.
This is a simplification, and in particular neglects the possibility that the shear shifts from being solar to anti-solar, but it allows us to approximately infer $c_2$ on average across systems.
Moreover, it allows us to infer $\beta$ in a way that accounts for the average effect of latitudinal differential rotation.
Note that with this parameterization $\lambda < 0$ corresponds to solar-type latitudinal differential rotation and $\lambda > 0$ to anti-solar differential rotation.

We take $\omega$ to be known, as the errors on the binary periods are tiny.
We take $\Omega_{\rm s}$ to be log-normally distributed with variance $\sigma$, which is a model parameter we assume to be universal across all systems in our sample.
Our likelihood function is then just the log-normal likelihood distribution of $\Omega_{\rm s}/\omega = P_{\rm orb}/P_{\rm s}$, and our model is equation~\eqref{eq:omegas} combined with equation~\eqref{eq:dlnO2}.

We exclude scenarios with extremely strong latitudinal differential rotation, specifically $c_2 < -8/5$ or $c_2 > 7/2$, because in these scenarios equation~\eqref{eq:omegas} predicts $P_{\rm orb}/P_{\rm s}<0$, which seems highly unlikely.
Our prior is uniform in $\beta$ on $[-3,3]$ because there are neither theoretical expectations nor observational indications of stronger radial shear than the extremes of this window.
Similarly, our prior is uniform in $\gamma$ on $[-2,2]$ because this encompasses all observational indications of which we are aware, and we are not aware of theoretical reasons for a more extreme radial variation of the latitudinal shear.
Our prior is uniform in $\lambda$ on $[-1,1]$ because latitudinal shear greater than unity is not expected nor has this been observed to our knowledge.
Finally, we take our prior over $\sigma$ to be uniform in $[0,1]$ because the uncertainty in a period measurement cannot be negative, and $\sigma=1$ corresponds to $100$~per-cent uncertainty, which is much greater than the reported observational uncertainties where available.
As we shall see none of our parameters exhibit significant posterior probability mass near the boundaries of these prior windows except where such boundaries are logically necessary (e.g. $\sigma \geq 0$) or where the parameter is very weakly constrained, so it is unlikely that the data favor a region of parameter space outside of our prior.

To study the dependence of differential rotation on rotation rate, as well as to cleanly separate the systems which are likely synchronous from those which may not be, we inferred the posterior distribution for $(\lambda,\beta,\gamma,\sigma)$ on five different period ranges:
\begin{itemize}
	\item $P_{\rm orb} \in [0,50]{\rm d}$
	\item $P_{\rm orb} \in [0,2]{\rm d}$
	\item $P_{\rm orb} \in [2,6]{\rm d}$
	\item $P_{\rm orb} \in [6,10]{\rm d}$
	\item $P_{\rm orb} \in [10,50]{\rm d}$
\end{itemize}
This allows us to explore, for instance, if the preferred sign of $\lambda$ is different for different period windows.
For each of these we performed the inference twice.
The first time was as described above.
The second time we accounted for the possibility of outliers or systems which have been misclassified by assigning each system a prior probability of being an outlier and hence not subject to our model.
This probability was set to $10^{-6}$, which places an effective floor of $10^{-6}$ on the likelihood of individual observations.
Finally, to check for trends with stellar mass we also performed each of these inferences on two subsets of the data: those with $M \leq 0.9 M_\odot$ and those with $M > 0.9 M_\odot$.

Data were processed in the \code{Python} programming language~\citep{Rossum:1995:PRM:869369} with the \code{NumPy} software package~\citep{book}
Posterior sampling was done using the Nested Sampling algorithm~\citep{2017arXiv170403459H} as implemented in the \code{dynesty} software package~\citep{2019arXiv190402180S}.
Figures were made with the \code{Matplotlib}~\citep{Hunter:2007} and \code{Corner}~\citep{2016JOSS....1...24F} plotting packages.

\section{Results}
\label{sec:res}

\begin{table*}
\label{tab:res}
\begin{tabular}{l | l | l | l | l | l | l | l | l | l | l | l | l | l | l }
	$P_{\rm orb, min}$ & $P_{\rm orb, max}$&& $\beta$ & $\beta_{1-\sigma}$ & $\beta_{2-\sigma}$ & $\lambda$ & $\lambda_{1-\sigma}$ & $\lambda_{2-\sigma}$ & $\gamma$ & $\gamma_{1-\sigma}$ & $\gamma_{2-\sigma}$ & $\sigma$ & $\sigma_{1-\sigma}$ & $\sigma_{2-\sigma}$\\
	&&Outliers& \multicolumn{3}{l}{Radial Shear} & \multicolumn{3}{l}{Latitudinal Shear} & \multicolumn{3}{l}{Latitudinal Shear} & \multicolumn{3}{l}{Rotation Period}\\
	&&Excluded?& &&& &&& \multicolumn{3}{l}{Period Scaling} & \multicolumn{3}{l}{Uncertainty}\\\hline

$0$ & $50$ & No & $-0.472$ & $^{-0.292}_{-0.736}$ & $^{0.037}_{-1.156}$ & $0.047$ & $^{0.115}_{-0.000}$ & $^{0.197}_{-0.069}$ & $0.573$ & $^{1.408}_{0.004}$ & $^{1.891}_{-0.481}$ & $0.325$ & $^{0.339}_{0.313}$ & $^{0.352}_{0.302}$\\ 
\rule{0pt}{15pt}$0$ & $2$ & No & $-0.425$ & $^{0.006}_{-0.850}$ & $^{1.094}_{-1.679}$ & $0.041$ & $^{0.602}_{-0.487}$ & $^{0.932}_{-0.874}$ & $1.088$ & $^{1.697}_{0.384}$ & $^{1.954}_{-0.170}$ & $0.477$ & $^{0.516}_{0.442}$ & $^{0.559}_{0.412}$\\ 
\rule{0pt}{15pt}$2$ & $6$ & No & $-0.424$ & $^{-0.059}_{-0.947}$ & $^{0.282}_{-1.538}$ & $0.090$ & $^{0.244}_{-0.048}$ & $^{0.424}_{-0.274}$ & $0.659$ & $^{1.578}_{-0.174}$ & $^{1.949}_{-0.826}$ & $0.258$ & $^{0.274}_{0.244}$ & $^{0.291}_{0.230}$\\ 
\rule{0pt}{15pt}$6$ & $10$ & No & $-1.223$ & $^{-0.605}_{-1.869}$ & $^{0.071}_{-2.500}$ & $0.114$ & $^{0.276}_{-0.014}$ & $^{0.433}_{-0.165}$ & $0.181$ & $^{1.208}_{-0.854}$ & $^{1.843}_{-1.744}$ & $0.200$ & $^{0.220}_{0.182}$ & $^{0.243}_{0.167}$\\ 
\rule{0pt}{15pt}$10$ & $50$ & No & $0.148$ & $^{1.274}_{-1.899}$ & $^{2.141}_{-2.786}$ & $-0.122$ & $^{0.278}_{-0.420}$ & $^{0.502}_{-0.585}$ & $-0.395$ & $^{0.912}_{-1.531}$ & $^{1.461}_{-1.913}$ & $0.312$ & $^{0.356}_{0.276}$ & $^{0.404}_{0.249}$\\ \hline 
\rule{0pt}{15pt}$0$ & $50$ & Yes & $0.152$ & $^{0.251}_{-0.482}$ & $^{0.337}_{-0.632}$ & $-0.105$ & $^{0.031}_{-0.126}$ & $^{0.056}_{-0.147}$ & $0.428$ & $^{0.525}_{-0.454}$ & $^{0.629}_{-0.643}$ & $0.082$ & $^{0.085}_{0.078}$ & $^{0.089}_{0.075}$\\ 
\rule{0pt}{15pt}$0$ & $2$ & Yes & $0.000$ & $^{0.010}_{-0.010}$ & $^{0.019}_{-0.022}$ & $0.079$ & $^{0.167}_{0.030}$ & $^{0.273}_{-0.004}$ & $1.571$ & $^{1.893}_{1.080}$ & $^{1.987}_{0.550}$ & $0.008$ & $^{0.009}_{0.008}$ & $^{0.010}_{0.007}$\\ 
\rule{0pt}{15pt}$2$ & $6$ & Yes & $0.031$ & $^{0.091}_{-0.252}$ & $^{0.139}_{-0.346}$ & $-0.079$ & $^{0.014}_{-0.138}$ & $^{0.028}_{-0.184}$ & $1.439$ & $^{1.884}_{-1.094}$ & $^{1.981}_{-1.503}$ & $0.038$ & $^{0.041}_{0.036}$ & $^{0.043}_{0.034}$\\ 
\rule{0pt}{15pt}$6$ & $10$ & Yes & $0.066$ & $^{0.373}_{-0.330}$ & $^{0.677}_{-1.186}$ & $-0.205$ & $^{-0.090}_{-0.287}$ & $^{0.084}_{-0.359}$ & $1.570$ & $^{1.878}_{0.884}$ & $^{1.983}_{-1.557}$ & $0.098$ & $^{0.109}_{0.090}$ & $^{0.121}_{0.082}$\\ 
\rule{0pt}{15pt}$10$ & $50$ & Yes & $0.264$ & $^{1.329}_{-1.911}$ & $^{2.267}_{-2.778}$ & $-0.161$ & $^{0.263}_{-0.429}$ & $^{0.498}_{-0.596}$ & $-0.495$ & $^{0.932}_{-1.549}$ & $^{1.508}_{-1.929}$ & $0.311$ & $^{0.352}_{0.277}$ & $^{0.400}_{0.249}$\\ 

	\hline	
\end{tabular}
\caption{Results are shown for the inference procedures described in Section~\ref{sec:infer}, excluding the mass subset procedure. Values of inferred parameters are posterior medians. Confidence intervals are $68$~per-cent ($1-\sigma$) and $95$~per-cent ($2-\sigma$) bounds. 
The full table including the mass subset results is available in the Supplementary Information and at~\url{Zenodo.org}}
\end{table*}

The results of our inference procedure are summarized in Table~\ref{tab:res}.
These exclude the inferences done on the subsets of our data with $M \leq 0.9 M_\odot$ and $M > 0.9 M_\odot$ because most of these did not show any significant differences in posterior distributions from one another.
The cases which do show a difference are discussed Section~\ref{sec:mass} and shown in Table~\ref{tab:res_mass}.

An immediate conclusion from Table~\ref{tab:res} is that there are indeed outliers in our sample.
This follows because the models which exclude outliers favor much smaller $\sigma$ values than those which do not, and that at short periods the former typically favor rotation period errors consistent with the $\sim 7$~per-cent reported by~\citet{2017A&A...605A.111C} from measurements of star spots on giants.

\subsection{Outliers}

\begin{figure}
	\includegraphics[width=0.5\textwidth]{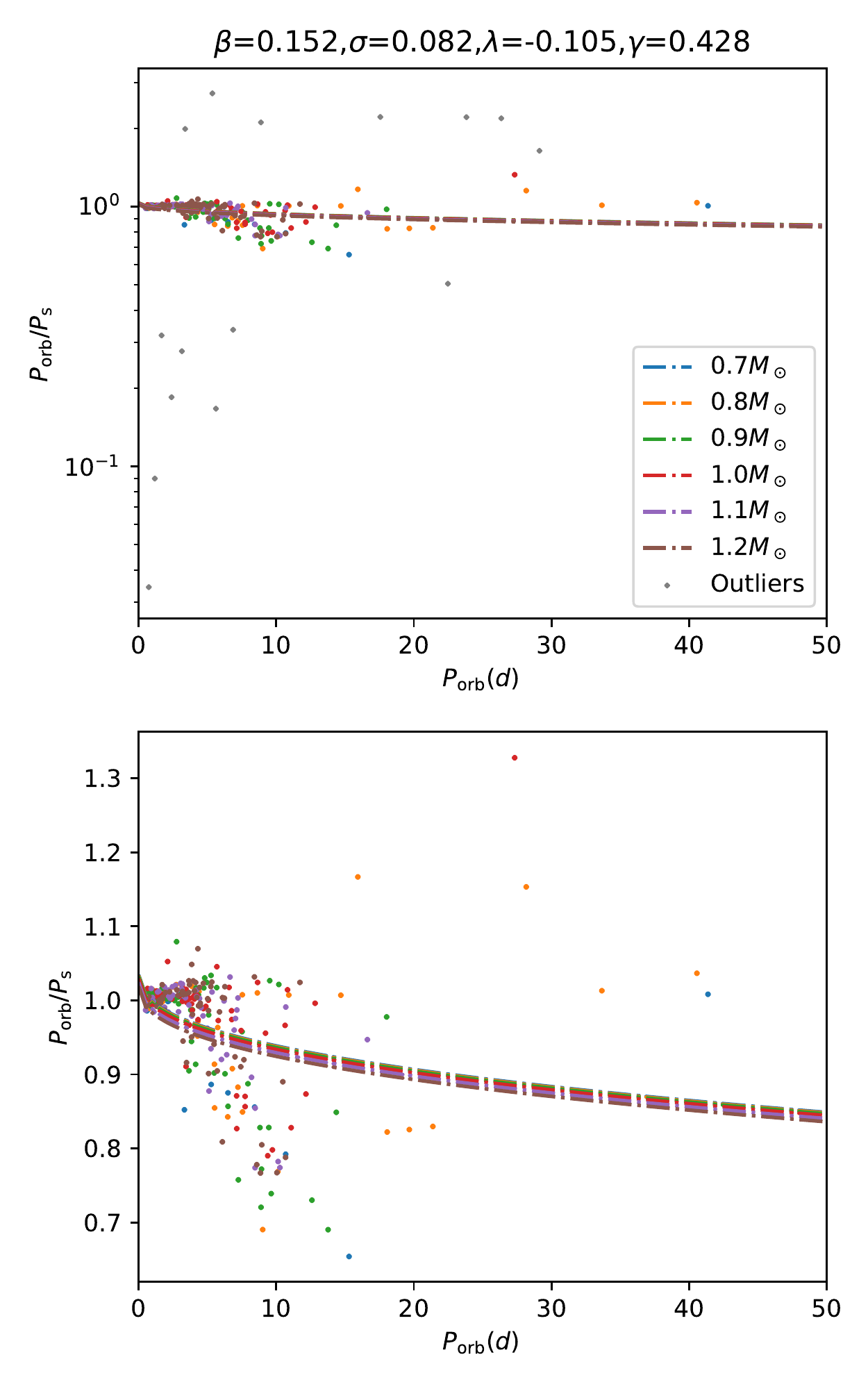}
	\caption{The ratio of orbital period to rotation period is shown as a function of the orbital period and stellar mass, indicated by color. Outliers were identified by the inference procedure as systems with a likelihood at the posterior median parameter values below $10^{-6}$, and are shown as grey circles. The upper panel shows all objects while the lower excludes outliers. These results are from the inference including systems with orbital periods less than $50\,{\rm d}$. The dashed lines indicate the model predictions at the median parameter values for the stellar masses of the same color. We note that many of the longer period systems are unlikely to be synchronized, accounting for the large differences between our model and the data at long periods.}
	\label{fig:outlier}
\end{figure}

To better understand which objects our methods label as outliers consider Figure~\ref{fig:outlier}, which shows our analysis of all systems with periods less than $50\,{\rm d}$.
The systems automatically identified as outliers are shown in gray.
The bulk of the sample is not labeled as outliers, and the systems which are those with either strongly super-synchronous rotation or the handful at strongly sub-synchronous rotation.
That is, there seems to be a clear distinction between the bulk of the systems and the outliers, which suggests that the outliers do not reflect the same population as the bulk.

To test our outlier identification we compared the outliers identified in Figure~\ref{fig:outlier} with that identified by a fit on those systems with orbital periods between $2\,{\rm d}$ and $6\,{\rm d}$.
Both the analysis of the full sample and the restricted subset identified the same outliers, suggesting that the outlier identification is robust to the details of the data.

We now focus on just the fits which exclude outliers.
The results of the remaining fits are presented in Appendix~\ref{appen:yes_outliers}.

\subsection{$P_{\rm orb} < 2\,{\rm d}$}

\begin{figure}
	\includegraphics[width=0.5\textwidth]{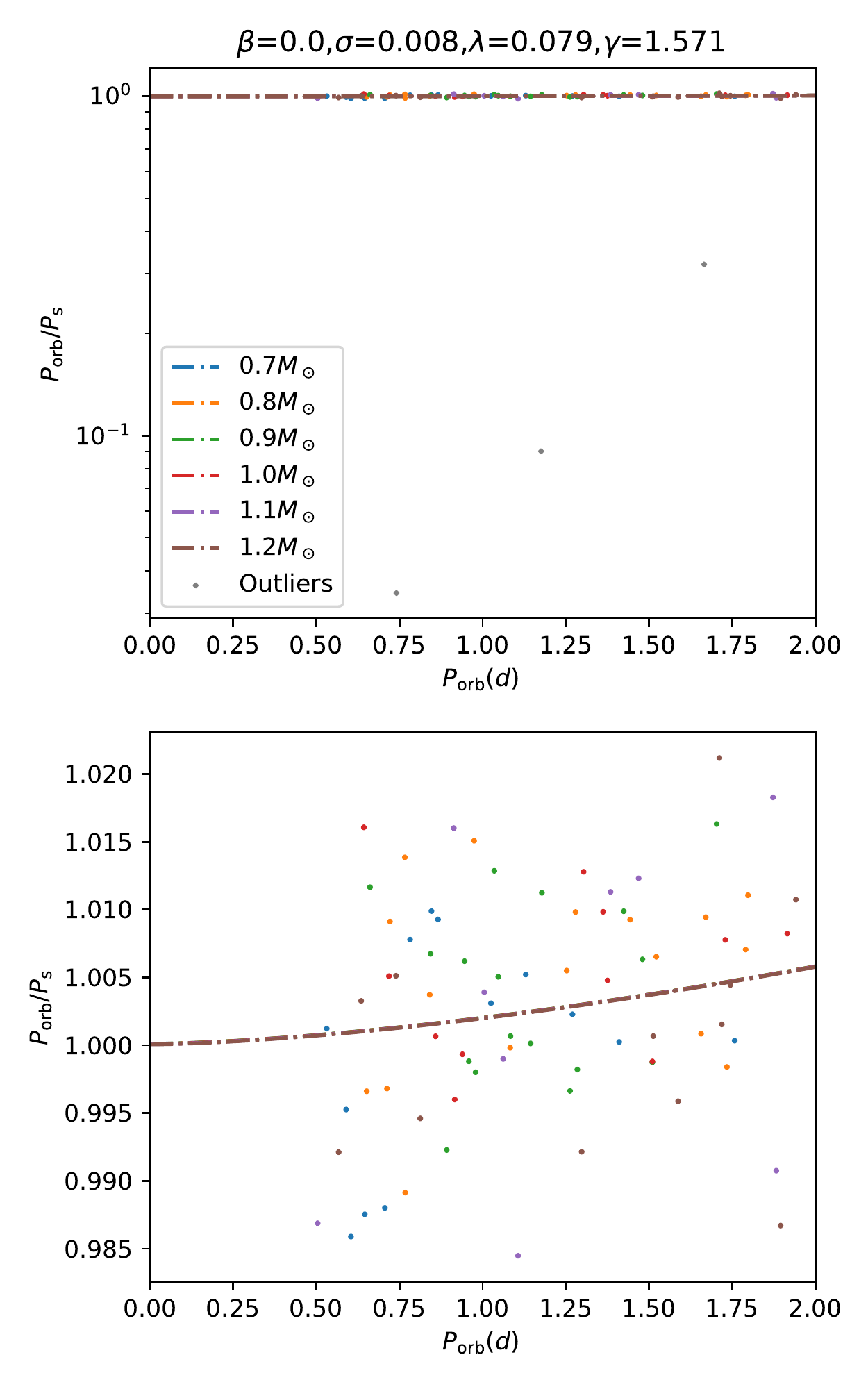}
	\caption{The ratio of orbital period to rotation period is shown as a function of the orbital period and stellar mass, indicated by color. Outliers were identified by the inference procedure as systems with a likelihood at the posterior median parameter values below $10^{-6}$, and are shown as grey circles. The upper panel shows all objects while the lower excludes outliers. These results are from the inference including systems with orbital periods less than $2\,{\rm d}$. The dashed lines indicate the model predictions at the median parameter values for the stellar masses of the same color.}
	\label{fig:02d}
\end{figure}

Figure~\ref{fig:02d} shows the ratio of $P_{\rm orb}$ to $P_{\rm s}$ as a function of $P_{\rm orb}$ for those systems with $P_{\rm orb} < 2\,{\rm d}$, along with our model for the median parameters fit to these systems.
There are three outliers with significantly sub-synchronous rotation.
Neglecting those the data are tightly clustered about the fit line.
Note the scale of the vertical axis: these systems all have orbital periodss within $2.5$~per-cent of their rotation periods.
Compared with the large range of period ratios in Figure~\ref{fig:kstar} this is an extremely narrow window around what we expect for systems without differential rotation.
This sharp distribution allows us to place significant constraints on the shear present in these systems.

\begin{figure}
	\includegraphics[width=0.5\textwidth]{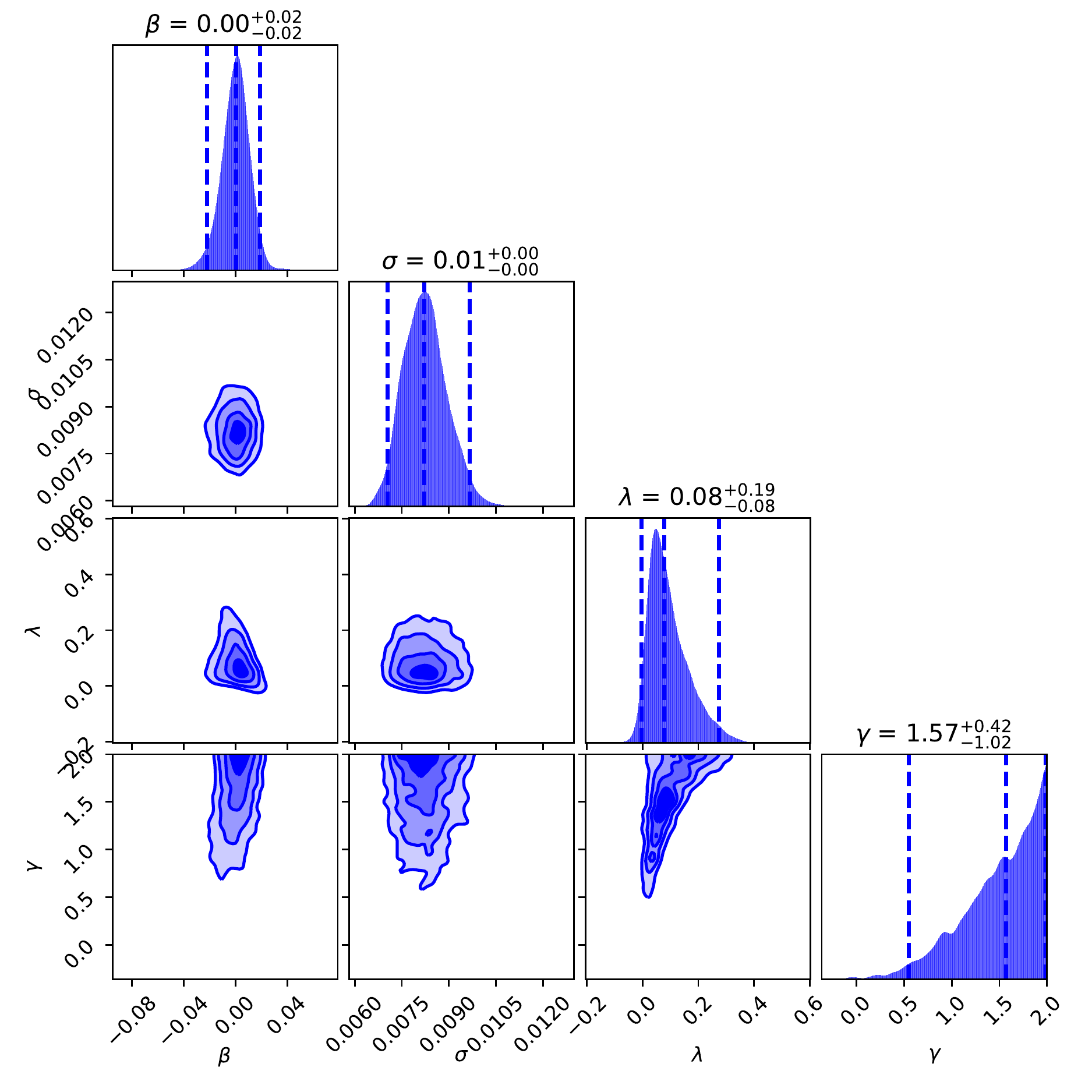}
	\caption{The marginalized posterior distribution over our model parameters is shown for the inference including systems with orbital periods less than $2\,{\rm d}$.}
	\label{fig:02d_post}
\end{figure}

The 2D marginalized posterior distribution is shown in Figure~\ref{fig:02d_post}.
The preferred uncertainty ($\sigma$) on the data is low, suggesting that the data are consistent with the model, have minimal intrinsic scatter, and have small observational uncertainties.

The distribution of the radial shear $\beta$ is tightly-peaked near $\beta=0$, corresponding to no or minimal radial differential rotation.
The preferred latitudinal differential rotation is small and anti-solar, with $\lambda > 0$.
This is in tension with what is seen in global 3D hydrodynamical convection simulations~\citep{2013ApJ...779..176G,2014MNRAS.438L..76G}, where at such short periods the rotation becomes solar-like.

Note that the scaling of this with period, parameterized by $\gamma$, is consistent with the value of $\gamma \approx 1.5$ seen by~\citet{2017AJ....154..250L}, though the uncertainties are large.
Nonetheless, at $2-\sigma$ confidence we find $0.5 < \gamma < 2$, indicating that latitudinal differential rotation increases with increasing orbital, and hence spin, period.

\subsection{$P_{\rm orb} > 2\,{\rm d}$}

For the most part the longer period windows show similar results, so we just summarize them here.
The detailed fits and marginalized posterior distributions for each of these fits are provided in Appendix~\ref{appen:full}.

As the orbital period increases the intrinsic scatter in the data, as measured by $\sigma$, increases.
For orbital periods greater than $10\,{\rm d}$ the scatter significantly exceeds typical reported observational uncertainties.
At a minimum this suggests that there may be additional intrinsic scatter in that population beyond what our model captures.
This could be due in part to the fact that at longer orbital periods systems are less likely to be synchronized.
Because our predictions do not apply to such systems the results for periods above $10\,{\rm d}$ should be taken with due caution, and we focus our interpretation on the shorter-period systems.

Following the trend of increasing scatter with increasing orbital period, the posterior distributions of our model parameters widen as we move to longer periods.
Despite this the median values remain quite similar.
We find that even at $2-\sigma$ the radial shear is small, with even our widest window running from $\beta=-1$ to $\beta=0.6$.

The latitudinal shear favors a small amount of solar differential rotation ($\lambda < 0$) at longer periods, though in all cases the $2-\sigma$ confidence intervals allow for anti-solar rotation as well.
In actuality, the posterior distribution is bimodal, with one solar mode and one anti-solar mode.
This is because at longer periods the data favor faster orbits than surface rotation ($P_{\rm orb} < P_{\rm s}$).
Because tides are sensitive to higher latitudes than the star spot measurements that can be achieved either by having the equator rotate slower than the poles ($\lambda > 0$) or by having the interior rotate faster than the surface ($\beta > 0$).

As before, the longer-period systems favor increasing latitudinal shear with increasing orbital period, though the uncertainties on the exponent $\gamma$ are large.

\subsection{Mass Dependence}
\label{sec:mass}

As stellar mass increases the convective envelope becomes shallower and hotter, and so the convective turnover becomes more rapid. This increases the Rossby number, which is thought to result in stronger relative differential rotation~\citep{2014MNRAS.438L..76G}. Hence, we should expect higher-mass stars at the same rotation period to have more shear than lower-mass stars.  

We see some tentative evidence for this trend, although any conclusive statements are limited by the size of our uncertainties. We show in Table~\ref{tab:res_mass}, the results of breaking our sample into two mass bins for periods less than $10\,{\rm d}$. For the high-mass sample we see that $\beta$ is consistent with zero in the $0-2\,{\rm d}$ period window and that $\beta > 0$ at nearly $2-\sigma$ confidence for both the $2-6\,{\rm d}$ and $6-10\,{\rm d}$ period windows. The transition around a period of $2\,{\rm d}$ is on the short end relative to typical convective turnover times of $10\,{\rm d}$ for these stars, but this is plausibly consistent given the theoretical uncertainties. 

We cannot identify a similar trend in the low mass sample, however, since the uncertainties in each period bin are large, and consistent with either following the high mass sample or having zero radial shear. If the trend in the higher-mass stars is real then the lower-mass ones presumably undergo a similar transition to beta greater than zero, but at a longer orbital period. At longer periods our sample is likely not synchronized, however, which may explain why we do not see such a transition.

\begin{table*}
\label{tab:res_mass}
\begin{tabular}{l | l | l | l | l | l | l | l | l | l | l | l | l | l | l}
	$P_{\rm orb, min}$ & $P_{\rm orb, max}$& Mass&$\beta$ & $\beta_{1-\sigma}$ & $\beta_{2-\sigma}$ & $\lambda$ & $\lambda_{1-\sigma}$ & $\lambda_{2-\sigma}$ & $\gamma$ & $\gamma_{1-\sigma}$ & $\gamma_{2-\sigma}$ & $\sigma$ & $\sigma_{1-\sigma}$ & $\sigma_{2-\sigma}$\\
	&& Selection& \multicolumn{3}{l}{Radial Shear} & \multicolumn{3}{l}{Latitudinal Shear} & \multicolumn{3}{l}{Latitudinal Shear} & \multicolumn{3}{l}{Rotation Period}\\
	&&& &&& &&& \multicolumn{3}{l}{Period Scaling} & \multicolumn{3}{l}{Uncertainty}\\\hline

\rule{0pt}{15pt}$0$ & $2$ & $\leq 0.9 M_\odot$ &$-0.014$ & $^{0.001}_{-0.067}$ & $^{0.014}_{-0.231}$ & $0.092$ & $^{0.211}_{0.044}$ & $^{0.368}_{0.011}$ & $1.270$ & $^{1.816}_{0.396}$ & $^{1.971}_{0.070}$ & $0.007$ & $^{0.008}_{0.007}$ & $^{0.009}_{0.006}$\\ 
\rule{0pt}{15pt}$0$ & $2$ & $> 0.9 M_\odot$ &$0.007$ & $^{0.031}_{-0.016}$ & $^{0.086}_{-0.041}$ & $0.041$ & $^{0.146}_{-0.017}$ & $^{0.289}_{-0.107}$ & $1.573$ & $^{1.894}_{0.933}$ & $^{1.986}_{-0.024}$ & $0.010$ & $^{0.011}_{0.009}$ & $^{0.013}_{0.008}$\\ 
\rule{0pt}{15pt}$2$ & $6$ & $\leq 0.9 M_\odot$ &$-1.291$ & $^{0.141}_{-2.262}$ & $^{0.409}_{-2.797}$ & $0.236$ & $^{0.490}_{-0.182}$ & $^{0.634}_{-0.286}$ & $-0.149$ & $^{1.253}_{-0.275}$ & $^{1.895}_{-0.561}$ & $0.044$ & $^{0.049}_{0.040}$ & $^{0.055}_{0.036}$\\ 
\rule{0pt}{15pt}$2$ & $6$ & $> 0.9 M_\odot$ &$0.109$ & $^{0.173}_{0.050}$ & $^{0.246}_{-0.010}$ & $-0.097$ & $^{-0.057}_{-0.142}$ & $^{-0.013}_{-0.190}$ & $1.551$ & $^{1.869}_{1.023}$ & $^{1.980}_{0.450}$ & $0.033$ & $^{0.036}_{0.030}$ & $^{0.039}_{0.028}$\\ 
\rule{0pt}{15pt}$6$ & $10$ & $\leq 0.9 M_\odot$ &$-0.261$ & $^{1.081}_{-1.615}$ & $^{2.541}_{-2.698}$ & $-0.122$ & $^{0.187}_{-0.444}$ & $^{0.469}_{-0.666}$ & $0.309$ & $^{1.298}_{-0.781}$ & $^{1.864}_{-1.670}$ & $0.117$ & $^{0.139}_{0.099}$ & $^{0.169}_{0.086}$\\ 
\rule{0pt}{15pt}$6$ & $10$ & $> 0.9 M_\odot$ &$0.684$ & $^{1.082}_{0.309}$ & $^{1.564}_{-0.060}$ & $-0.315$ & $^{-0.228}_{-0.399}$ & $^{-0.127}_{-0.477}$ & $1.508$ & $^{1.834}_{1.054}$ & $^{1.971}_{0.559}$ & $0.084$ & $^{0.097}_{0.074}$ & $^{0.112}_{0.067}$\\ 

	\hline	
\end{tabular}
\caption{Results are shown for the inference procedures described in Section~\ref{sec:infer} with outliers excluded and periods restricted to less than $10\,{\rm d}$. Values of inferred parameters are posterior medians. Confidence intervals are $68$~per-cent ($1-\sigma$) and $95$~per-cent ($2-\sigma$) bounds.}
\end{table*}

\section{Discussion}
\label{sec:disc}

Differential rotation in binary star systems creates a difference between the surface rotation rate and the orbital period.
We have used this difference along with a parameterized rotation model to infer the radial and latitudinal shear in main-sequence eclipsing binary systems.
Our principle finding is that, for orbital periods less than $10\,{\rm d}$, main sequence K-F stars exhibit little radial or latitudinal shear in their convection zones.
Consistent with theoretical expectations, we see tentative evidence of this shear increasing with stellar mass but remaining small in absolute terms.

In our model, the latitudinal shear is given by
\begin{align}
	|\partial_\theta \ln\Omega| \approx c_2 \approx \lambda \left(\frac{P_{\rm s}}{10{\rm d}}\right)^{\gamma},
\end{align}
where our preferred values for $\lambda$ range from $-0.2$ to $0.1$.
The variation of this shear with orbital period is parameterized by $\gamma$.
For periods less than $2\,{\rm d}$ this is positive and lies between $1-2$, which is consistent with the findings of~\citet{2017AJ....154..250L}.
For longer periods the constraints become weaker and very little can be said about the dependence of latitudinal shear on period.

An interesting comparison is provided by~\citet{2018Sci...361.1231B}, who searched for asteroseismic signatures of latitudinal differential rotation in 40 solar-type stars and found evidence of non-zero shear in 13 of them.
Of these, 5 have periods below $10\,{\rm d}$, so we compare with those.
Because none of these objects have periods less than $6\,{\rm d}$, the relevant comparison is to our results between $6\,{\rm d}$ and $10\,{\rm d}$, where we find $\lambda$ ranging from $-0.09$ to $-0.29$ and $\gamma$ between $0.88$ and $1.88$, giving $c_2$ between $-0.03$ and $-0.29$.

\citet{2018Sci...361.1231B} find shear between the pole and the equator of between $-0.7$ and $-2.1$ times the equatorial rotation rate.
We can relate this measure to $c_2$ through equation~\eqref{eq:rot_prof}, which yields
\begin{align}
	\frac{\Omega_{\rm pole}-\Omega_{\rm equator}}{\Omega_{\rm equator}} = \frac{3 c_2}{2 - c_2},
\end{align}
which means that their range of shears corresponds to $c_2$ ranging from $-0.61$ to $-4.7$ with an unweighted average of $-2.1$.
These shears are much larger than anything we find.
Even if we average them with the remaining two-thirds of the sample with no significant detection of shear we obtain a value of order $-0.7$, which is larger than what we obtained in this work.

There are a few possible explanations for this discrepancy.
One is that starspots are preferentially at lower latitudes than we have assumed.
That would cause us to under-estimate the latitudinal shear.
It is also the case that our methods are sensitive to different regions of the star than theirs because the asteroseismic kernel does not have the same form as equation~\eqref{eq:kstar}.
Thus it could be that the large shears they detect occur in a different part of the star than the regions our method probes.
In either case this is clearly a discrepancy which merits further investigation.

For the radial shear we find that
\begin{align}
	|r\partial_r \ln \Omega| \approx \beta,
\end{align}
For periods less than $2\,{\rm d}$ we find a particularly strong constraint of $|\beta| < 0.02$, which is somewhat smaller than the result we obtain by fitting our model rotation profile to that of the Sun (see Section~\ref{sec:profile}).
This could be a result of these stars having much more rapid rotation rates.
Three-dimensional MHD simulations of the convection zones of rapidly rotating solar-type stars show differential rotation which increases sub-linearly with rotation rate~\citep{2016csss.confE.152A,2017ApJ...836..192B}, such that more rapidly-rotating stars exhibit smaller $\beta$.
This is consistent with what we see.

Even for longer periods out to $10\,{\rm d}$ we find $\beta$ ranging from $0$ to $0.6$.
This is significant because~\citet{2015ApJ...808...35K} find that steep rotation profiles with $\beta < -1$ are needed to explain the rotation rates of red giant cores if the shear is primarily located in the convection zone.
For orbital periods, and hence rotation periods, less than $6\,{\rm d}$ we disfavor such steep profiles at $p < 10^{-4}$.
For periods less than $10\,{\rm d}$ we disfavor $\beta < -1$ with $p < 0.035$.
Even for periods greater than $10\,{\rm d}$ we disfavor steep slopes, though somewhat less strongly in part because we have fewer data points and so less inferential power.

Our data are mostly for main-sequence solar-type stars, with typical convective turnover times of $\tau \approx 10\,{\rm d}$.
If $\beta$ is only a function of the convective Rossby number~\citep{2013MNRAS.431.2200L,2015ApJ...808...35K}, then the spin periods for giants (which have longer convective turnover times) at the same Rossby number are longer.
In particular the Rossby number scales as
\begin{align}
	\mathrm{Ro} &\sim \frac{1}{\tau \Omega}\sim \left(\frac{R_\star}{h}\right)\frac{u_c}{R_\star \Omega}.
\end{align}
The convective velocity is related to the heat flux $F$ by
\begin{align}
	F \approx \rho u_c^3,
\end{align}
so
\begin{align}
	\mathrm{Ro}&\sim \left(\frac{R_\star}{h}\right)\left(\frac{F}{\rho R_\star^3 \Omega^3}\right)^{1/3}\sim \left(\frac{R_\star}{h}\right)\left(\frac{T_{\rm eff}^4}{M \Omega^3}\right)^{1/3},
\end{align}
where $M$ is the mass of the star.
So to hold $\mathrm{Ro}$ fixed,
\begin{align}
	P_{\rm s} \propto \frac{M^{1/3}}{T_{\rm eff}^{4/3}}\left(\frac{h}{R_\star}\right).
\end{align}
At fixed mass this becomes
\begin{align}
	P_{\rm s} \propto T_{\rm eff}^{-4/3}\left(\frac{h}{R_\star}\right).
\end{align}
Moving from a solar model where $h/R_\star \sim 0.1$ and $T_{\rm eff} \sim T_{\odot}$ to a giant where $h/R_\star \sim 1$ and $T_{\rm eff} \sim 0.7 T_{\odot}$ then increases the period by roughly$15$-fold.

Hence, for giants with $P_{\rm s} < 75\,{\rm d}$, the Rossby numbers are similar to those in the solar type stars studied here, and we thus suggest that there is little radial shear in their convection zones and that the bulk of the core-envelope differential rotation likely lies in their radiative zones.
This is in agreement with the asteroseismic constraints of~\citet{2017MNRAS.464L..16K}, who find that $\beta > 1$ is inconsistent with the rotational splittings for Kepler-56.

While more data on solar-type stars would be useful in tightening our constraints on both the radial and latitudinal shears, data on red giants may prove even more valuable.
Figure~\ref{fig:kstar} shows that $k_\star(\beta)$, which controls the size of the signal we expect, is much larger for red giants than for main-sequence solar-type stars.
This is because the former have much deeper convection zones.
Hence, stronger constraints may come from such systems even if there are fewer of them.
Moreover, for some such systems it may be possible to obtain core rotation rates for these from astereoseismology~\citep[see, e.g.][]{2012Natur.481...55B}.
With simultaneous core and surface rotation data as well as tidal constraints on shear in the cores of these stars it would be possible to estimate the shear present in the radiative zones, which would provide a valuable test of various theoretical angular momentum transport mechanisms~\citep{2002A&A...381..923S,2003A&A...411..543M,2014ApJ...788...93C,2019MNRAS.485.3661F,2019NatAs.tmp..320B}.

\section*{Acknowledgements}

ASJ thanks the Gordon and Betty Moore Foundation (Grant GBMF7392) and the National Science Foundation (Grant No. NSF PHY-1748958) for supporting this work.
JT acknowledges that support for this work was provided by NASA through the NASA Hubble Fellowship grant No. 51424 awarded by the Space Telescope Science Institute, which is operated by the Association of Universities for Research in Astronomy, Inc., for NASA, under contract NAS5-26555
This research was partially conducted during the Exostar19 program at the Kavli Institute for Theoretical Physics at UC Santa Barbara, which was supported in part by the National Science Foundation under Grant No. NSF PHY-1748958.
This work was supported by the Flatiron Institute of the Simons Foundation.
The authors also thank Lyra Cao for catching errors in an earlier version of this manuscript.

%%%%%%%%%%%%%%%%%%%%%%%%%%%%%%%%%%%%%%%%%%%%%%%%%%



\bibliographystyle{mnras}
\bibliography{refs}

\begin{thebibliography}{}
\makeatletter
\relax
\def\mn@urlcharsother{\let\do\@makeother \do\$\do\&\do\#\do\^\do\_\do\%\do\~}
\def\mn@doi{\begingroup\mn@urlcharsother \@ifnextchar [ {\mn@doi@}
  {\mn@doi@[]}}
\def\mn@doi@[#1]#2{\def\@tempa{#1}\ifx\@tempa\@empty \href
  {http://dx.doi.org/#2} {doi:#2}\else \href {http://dx.doi.org/#2} {#1}\fi
  \endgroup}
\def\mn@eprint#1#2{\mn@eprint@#1:#2::\@nil}
\def\mn@eprint@arXiv#1{\href {http://arxiv.org/abs/#1} {{\tt arXiv:#1}}}
\def\mn@eprint@dblp#1{\href {http://dblp.uni-trier.de/rec/bibtex/#1.xml}
  {dblp:#1}}
\def\mn@eprint@#1:#2:#3:#4\@nil{\def\@tempa {#1}\def\@tempb {#2}\def\@tempc
  {#3}\ifx \@tempc \@empty \let \@tempc \@tempb \let \@tempb \@tempa \fi \ifx
  \@tempb \@empty \def\@tempb {arXiv}\fi \@ifundefined
  {mn@eprint@\@tempb}{\@tempb:\@tempc}{\expandafter \expandafter \csname
  mn@eprint@\@tempb\endcsname \expandafter{\@tempc}}}

\bibitem[\protect\citeauthoryear{{Abdul-Masih} et~al.,}{{Abdul-Masih}
  et~al.}{2016}]{2016AJ....151..101A}
{Abdul-Masih} M.,  et~al., 2016, \mn@doi [\aj] {10.3847/0004-6256/151/4/101},
  \href {https://ui.adsabs.harvard.edu/abs/2016AJ....151..101A} {151, 101}

\bibitem[\protect\citeauthoryear{{Antia} \& {Basu}}{{Antia} \&
  {Basu}}{2010}]{2010ApJ...720..494A}
{Antia} H.~M.,  {Basu} S.,  2010, \mn@doi [\apj] {10.1088/0004-637X/720/1/494},
  \href {https://ui.adsabs.harvard.edu/abs/2010ApJ...720..494A} {720, 494}

\bibitem[\protect\citeauthoryear{Antia \& Chitre}{Antia \&
  Chitre}{2016}]{Private}
Antia H.~M.,  Chitre S.~M.,  private communication, 2016

\bibitem[\protect\citeauthoryear{{Antia}, {Basu}  \& {Chitre}}{{Antia}
  et~al.}{2008}]{2008ApJ...681..680A}
{Antia} H.~M.,  {Basu} S.,   {Chitre} S.~M.,  2008, \mn@doi [\apj]
  {10.1086/588523}, \href {http://adsabs.harvard.edu/abs/2008ApJ...681..680A}
  {681, 680}

\bibitem[\protect\citeauthoryear{{Augustson}, {Mathis}  \& {Brun}}{{Augustson}
  et~al.}{2016}]{2016csss.confE.152A}
{Augustson} K.,  {Mathis} S.,   {Brun} A.~S.,  2016, in 19th Cambridge Workshop
  on Cool Stars, Stellar Systems, and the Sun (CS19). p.~152,
  \mn@doi{10.5281/zenodo.237251}

\bibitem[\protect\citeauthoryear{{Beck} et~al.,}{{Beck}
  et~al.}{2012}]{2012Natur.481...55B}
{Beck} P.~G.,  et~al., 2012, \mn@doi [\nat] {10.1038/nature10612}, \href
  {https://ui.adsabs.harvard.edu/abs/2012Natur.481...55B} {481, 55}

\bibitem[\protect\citeauthoryear{{Benomar} et~al.,}{{Benomar}
  et~al.}{2018}]{2018Sci...361.1231B}
{Benomar} O.,  et~al., 2018, \mn@doi [Science] {10.1126/science.aao6571}, \href
  {https://ui.adsabs.harvard.edu/abs/2018Sci...361.1231B} {361, 1231}

\bibitem[\protect\citeauthoryear{{Borkovits}, {Hajdu}, {Sztakovics},
  {Rappaport}, {Levine}, {B{\'{\i}}r{\'o}}  \& {Klagyivik}}{{Borkovits}
  et~al.}{2016}]{2016MNRAS.455.4136B}
{Borkovits} T.,  {Hajdu} T.,  {Sztakovics} J.,  {Rappaport} S.,  {Levine} A.,
  {B{\'{\i}}r{\'o}} I.~B.,   {Klagyivik} P.,  2016, \mn@doi [\mnras]
  {10.1093/mnras/stv2530}, \href
  {https://ui.adsabs.harvard.edu/abs/2016MNRAS.455.4136B} {455, 4136}

\bibitem[\protect\citeauthoryear{{Bowman} et~al.,}{{Bowman}
  et~al.}{2019}]{2019NatAs.tmp..320B}
{Bowman} D.~M.,  et~al., 2019, \mn@doi [Nature Astronomy]
  {10.1038/s41550-019-0768-1}, \href
  {https://ui.adsabs.harvard.edu/abs/2019NatAs.tmp..320B} {}

\bibitem[\protect\citeauthoryear{{Brun} et~al.,}{{Brun}
  et~al.}{2017}]{2017ApJ...836..192B}
{Brun} A.~S.,  et~al., 2017, \mn@doi [\apj] {10.3847/1538-4357/aa5c40}, \href
  {http://adsabs.harvard.edu/abs/2017ApJ...836..192B} {836, 192}

\bibitem[\protect\citeauthoryear{{Buchler} \& {Yueh}}{{Buchler} \&
  {Yueh}}{1976}]{Buchler1976}
{Buchler} J.~R.,  {Yueh} W.~R.,  1976, \mn@doi [\apj] {10.1086/154847}, \href
  {http://adsabs.harvard.edu/abs/1976ApJ...210..440B} {210, 440}

\bibitem[\protect\citeauthoryear{{Cantiello}, {Mankovich}, {Bildsten},
  {Christensen-Dalsgaard}  \& {Paxton}}{{Cantiello}
  et~al.}{2014}]{2014ApJ...788...93C}
{Cantiello} M.,  {Mankovich} C.,  {Bildsten} L.,  {Christensen-Dalsgaard} J.,
  {Paxton} B.,  2014, \mn@doi [\apj] {10.1088/0004-637X/788/1/93}, \href
  {https://ui.adsabs.harvard.edu/abs/2014ApJ...788...93C} {788, 93}

\bibitem[\protect\citeauthoryear{{Cassisi}, {Potekhin}, {Pietrinferni},
  {Catelan}  \& {Salaris}}{{Cassisi} et~al.}{2007}]{Cassisi2007}
{Cassisi} S.,  {Potekhin} A.~Y.,  {Pietrinferni} A.,  {Catelan} M.,   {Salaris}
  M.,  2007, \mn@doi [\apj] {10.1086/516819}, \href
  {http://adsabs.harvard.edu/abs/2007ApJ...661.1094C} {661, 1094}

\bibitem[\protect\citeauthoryear{{Ceillier} et~al.,}{{Ceillier}
  et~al.}{2017}]{2017A&A...605A.111C}
{Ceillier} T.,  et~al., 2017, \mn@doi [\aap] {10.1051/0004-6361/201629884},
  \href {https://ui.adsabs.harvard.edu/abs/2017A%26A...605A.111C} {605, A111}

\bibitem[\protect\citeauthoryear{{Chugunov}, {Dewitt}  \&
  {Yakovlev}}{{Chugunov} et~al.}{2007}]{Chugunov2007}
{Chugunov} A.~I.,  {Dewitt} H.~E.,   {Yakovlev} D.~G.,  2007, \mn@doi [\prd]
  {10.1103/PhysRevD.76.025028}, \href
  {https://ui.adsabs.harvard.edu/abs/2007PhRvD..76b5028C} {76, 025028}

\bibitem[\protect\citeauthoryear{{Claret} \& {Torres}}{{Claret} \&
  {Torres}}{2017}]{2017ApJ...849...18C}
{Claret} A.,  {Torres} G.,  2017, \mn@doi [\apj] {10.3847/1538-4357/aa8770},
  \href {https://ui.adsabs.harvard.edu/abs/2017ApJ...849...18C} {849, 18}

\bibitem[\protect\citeauthoryear{{Conroy} et~al.,}{{Conroy}
  et~al.}{2014}]{2014PASP..126..914C}
{Conroy} K.~E.,  et~al., 2014, \mn@doi [\pasp] {10.1086/678953}, \href
  {https://ui.adsabs.harvard.edu/abs/2014PASP..126..914C} {126, 914}

\bibitem[\protect\citeauthoryear{{Cyburt} et~al.,}{{Cyburt}
  et~al.}{2010}]{Cyburt2010}
{Cyburt} R.~H.,  et~al., 2010, \mn@doi [\apjs] {10.1088/0067-0049/189/1/240},
  \href {http://adsabs.harvard.edu/abs/2010ApJS..189..240C} {189, 240}

\bibitem[\protect\citeauthoryear{{Eggenberger}, {Buldgen}  \&
  {Salmon}}{{Eggenberger} et~al.}{2019}]{2019A&A...626L...1E}
{Eggenberger} P.,  {Buldgen} G.,   {Salmon} S.~J.~A.~J.,  2019, \mn@doi [\aap]
  {10.1051/0004-6361/201935509}, \href
  {https://ui.adsabs.harvard.edu/abs/2019A%26A...626L...1E} {626, L1}

\bibitem[\protect\citeauthoryear{{Ferguson}, {Alexander}, {Allard}, {Barman},
  {Bodnarik}, {Hauschildt}, {Heffner-Wong}  \& {Tamanai}}{{Ferguson}
  et~al.}{2005}]{Ferguson2005}
{Ferguson} J.~W.,  {Alexander} D.~R.,  {Allard} F.,  {Barman} T.,  {Bodnarik}
  J.~G.,  {Hauschildt} P.~H.,  {Heffner-Wong} A.,   {Tamanai} A.,  2005,
  \mn@doi [\apj] {10.1086/428642}, \href
  {http://adsabs.harvard.edu/abs/2005ApJ...623..585F} {623, 585}

\bibitem[\protect\citeauthoryear{{Foreman-Mackey}}{{Foreman-Mackey}}{2016}]{2016JOSS....1...24F}
{Foreman-Mackey} D.,  2016, \mn@doi [The Journal of Open Source Software]
  {10.21105/joss.00024}, \href
  {https://ui.adsabs.harvard.edu/abs/2016JOSS....1...24F} {1}

\bibitem[\protect\citeauthoryear{{Fuller}, {Fowler}  \& {Newman}}{{Fuller}
  et~al.}{1985}]{Fuller1985}
{Fuller} G.~M.,  {Fowler} W.~A.,   {Newman} M.~J.,  1985, \mn@doi [\apj]
  {10.1086/163208}, \href {http://adsabs.harvard.edu/abs/1985ApJ...293....1F}
  {293, 1}

\bibitem[\protect\citeauthoryear{{Fuller}, {Piro}  \& {Jermyn}}{{Fuller}
  et~al.}{2019}]{2019MNRAS.485.3661F}
{Fuller} J.,  {Piro} A.~L.,   {Jermyn} A.~S.,  2019, \mn@doi [\mnras]
  {10.1093/mnras/stz514}, \href
  {https://ui.adsabs.harvard.edu/abs/2019MNRAS.485.3661F} {485, 3661}

\bibitem[\protect\citeauthoryear{{Gastine}, {Yadav}, {Morin}, {Reiners}  \&
  {Wicht}}{{Gastine} et~al.}{2014}]{2014MNRAS.438L..76G}
{Gastine} T.,  {Yadav} R.~K.,  {Morin} J.,  {Reiners} A.,   {Wicht} J.,  2014,
  \mn@doi [\mnras] {10.1093/mnrasl/slt162}, \href
  {https://ui.adsabs.harvard.edu/abs/2014MNRAS.438L..76G} {438, L76}

\bibitem[\protect\citeauthoryear{{Gaulme}, {Jackiewicz}, {Appourchaux}  \&
  {Mosser}}{{Gaulme} et~al.}{2014}]{2014ApJ...785....5G}
{Gaulme} P.,  {Jackiewicz} J.,  {Appourchaux} T.,   {Mosser} B.,  2014, \mn@doi
  [\apj] {10.1088/0004-637X/785/1/5}, \href
  {https://ui.adsabs.harvard.edu/abs/2014ApJ...785....5G} {785, 5}

\bibitem[\protect\citeauthoryear{{Goldreich} \& {Nicholson}}{{Goldreich} \&
  {Nicholson}}{1977}]{1977Icar...30..301G}
{Goldreich} P.,  {Nicholson} P.~D.,  1977, \mn@doi [\icarus]
  {10.1016/0019-1035(77)90163-4}, \href
  {https://ui.adsabs.harvard.edu/abs/1977Icar...30..301G} {30, 301}

\bibitem[\protect\citeauthoryear{{Goldreich} \& {Soter}}{{Goldreich} \&
  {Soter}}{1966}]{1966Icar....5..375G}
{Goldreich} P.,  {Soter} S.,  1966, \mn@doi [\icarus]
  {10.1016/0019-1035(66)90051-0}, \href
  {https://ui.adsabs.harvard.edu/abs/1966Icar....5..375G} {5, 375}

\bibitem[\protect\citeauthoryear{{Guerrero}, {Smolarkiewicz}, {Kosovichev}  \&
  {Mansour}}{{Guerrero} et~al.}{2013}]{2013ApJ...779..176G}
{Guerrero} G.,  {Smolarkiewicz} P.~K.,  {Kosovichev} A.~G.,   {Mansour} N.~N.,
  2013, \mn@doi [\apj] {10.1088/0004-637X/779/2/176}, \href
  {http://adsabs.harvard.edu/abs/2013ApJ...779..176G} {779, 176}

\bibitem[\protect\citeauthoryear{{Hermes} et~al.,}{{Hermes}
  et~al.}{2017}]{2017ApJS..232...23H}
{Hermes} J.~J.,  et~al., 2017, \mn@doi [\apjs] {10.3847/1538-4365/aa8bb5},
  \href {https://ui.adsabs.harvard.edu/abs/2017ApJS..232...23H} {232, 23}

\bibitem[\protect\citeauthoryear{{Higson}, {Handley}, {Hobson}  \&
  {Lasenby}}{{Higson} et~al.}{2017}]{2017arXiv170403459H}
{Higson} E.,  {Handley} W.,  {Hobson} M.,   {Lasenby} A.,  2017, arXiv
  e-prints, \href {https://ui.adsabs.harvard.edu/abs/2017arXiv170403459H} {}

\bibitem[\protect\citeauthoryear{{Hubbard}}{{Hubbard}}{1974}]{1974Icar...23...42H}
{Hubbard} W.~B.,  1974, \mn@doi [\icarus] {10.1016/0019-1035(74)90102-X}, \href
  {https://ui.adsabs.harvard.edu/abs/1974Icar...23...42H} {23, 42}

\bibitem[\protect\citeauthoryear{Hunter}{Hunter}{2007}]{Hunter:2007}
Hunter J.~D.,  2007, \mn@doi [Computing in Science \& Engineering]
  {10.1109/MCSE.2007.55}, 9, 90

\bibitem[\protect\citeauthoryear{{Iglesias} \& {Rogers}}{{Iglesias} \&
  {Rogers}}{1993}]{Iglesias1993}
{Iglesias} C.~A.,  {Rogers} F.~J.,  1993, \mn@doi [\apj] {10.1086/172958},
  \href {http://adsabs.harvard.edu/abs/1993ApJ...412..752I} {412, 752}

\bibitem[\protect\citeauthoryear{{Iglesias} \& {Rogers}}{{Iglesias} \&
  {Rogers}}{1996}]{Iglesias1996}
{Iglesias} C.~A.,  {Rogers} F.~J.,  1996, \mn@doi [\apj] {10.1086/177381},
  \href {http://adsabs.harvard.edu/abs/1996ApJ...464..943I} {464, 943}

\bibitem[\protect\citeauthoryear{{Itoh}, {Hayashi}, {Nishikawa}  \&
  {Kohyama}}{{Itoh} et~al.}{1996}]{Itoh1996}
{Itoh} N.,  {Hayashi} H.,  {Nishikawa} A.,   {Kohyama} Y.,  1996, \mn@doi
  [\apjs] {10.1086/192264}, \href
  {http://adsabs.harvard.edu/abs/1996ApJS..102..411I} {102, 411}

\bibitem[\protect\citeauthoryear{{Kissin} \& {Thompson}}{{Kissin} \&
  {Thompson}}{2015}]{2015ApJ...808...35K}
{Kissin} Y.,  {Thompson} C.,  2015, \mn@doi [\apj]
  {10.1088/0004-637X/808/1/35}, \href
  {https://ui.adsabs.harvard.edu/abs/2015ApJ...808...35K} {808, 35}

\bibitem[\protect\citeauthoryear{{Kitchatinov} \& {R{\"u}diger}}{{Kitchatinov}
  \& {R{\"u}diger}}{1999}]{1999A&A...344..911K}
{Kitchatinov} L.~L.,  {R{\"u}diger} G.,  1999, \aap, \href
  {https://ui.adsabs.harvard.edu/abs/1999A%26A...344..911K} {344, 911}

\bibitem[\protect\citeauthoryear{{Klion} \& {Quataert}}{{Klion} \&
  {Quataert}}{2017}]{2017MNRAS.464L..16K}
{Klion} H.,  {Quataert} E.,  2017, \mn@doi [\mnras] {10.1093/mnrasl/slw171},
  \href {https://ui.adsabs.harvard.edu/abs/2017MNRAS.464L..16K} {464, L16}

\bibitem[\protect\citeauthoryear{{LaCourse} et~al.,}{{LaCourse}
  et~al.}{2015}]{2015MNRAS.452.3561L}
{LaCourse} D.~M.,  et~al., 2015, \mn@doi [\mnras] {10.1093/mnras/stv1475},
  \href {https://ui.adsabs.harvard.edu/abs/2015MNRAS.452.3561L} {452, 3561}

\bibitem[\protect\citeauthoryear{{Langanke} \&
  {Mart{\'{\i}}nez-Pinedo}}{{Langanke} \&
  {Mart{\'{\i}}nez-Pinedo}}{2000}]{Langanke2000}
{Langanke} K.,  {Mart{\'{\i}}nez-Pinedo} G.,  2000, \mn@doi [Nuclear Physics A]
  {10.1016/S0375-9474(00)00131-7}, \href
  {http://adsabs.harvard.edu/abs/2000NuPhA.673..481L} {673, 481}

\bibitem[\protect\citeauthoryear{{Lesaffre}, {Chitre}, {Potter}  \&
  {Tout}}{{Lesaffre} et~al.}{2013}]{2013MNRAS.431.2200L}
{Lesaffre} P.,  {Chitre} S.~M.,  {Potter} A.~T.,   {Tout} C.~A.,  2013, \mn@doi
  [\mnras] {10.1093/mnras/stt317}, \href
  {https://ui.adsabs.harvard.edu/abs/2013MNRAS.431.2200L} {431, 2200}

\bibitem[\protect\citeauthoryear{Li, Van Reeth, Bedding, Murphy  \& Antoci}{Li
  et~al.}{2019}]{10.1093/mnras/stz1171}
Li G.,  Van Reeth T.,  Bedding T.~R.,  Murphy S.~J.,   Antoci V.,  2019,
  \mn@doi [Monthly Notices of the Royal Astronomical Society]
  {10.1093/mnras/stz1171}, 487, 782

\bibitem[\protect\citeauthoryear{{Lurie} et~al.,}{{Lurie}
  et~al.}{2017}]{2017AJ....154..250L}
{Lurie} J.~C.,  et~al., 2017, \mn@doi [\aj] {10.3847/1538-3881/aa974d}, \href
  {https://ui.adsabs.harvard.edu/abs/2017AJ....154..250L} {154, 250}

\bibitem[\protect\citeauthoryear{{Maeder} \& {Meynet}}{{Maeder} \&
  {Meynet}}{2003}]{2003A&A...411..543M}
{Maeder} A.,  {Meynet} G.,  2003, \mn@doi [\aap] {10.1051/0004-6361:20031491},
  \href {https://ui.adsabs.harvard.edu/abs/2003A&A...411..543M} {411, 543}

\bibitem[\protect\citeauthoryear{{Matijevi{\v c}}, {Pr{\v s}a}, {Orosz},
  {Welsh}, {Bloemen}  \& {Barclay}}{{Matijevi{\v c}}
  et~al.}{2012}]{2012AJ....143..123M}
{Matijevi{\v c}} G.,  {Pr{\v s}a} A.,  {Orosz} J.~A.,  {Welsh} W.~F.,
  {Bloemen} S.,   {Barclay} T.,  2012, \mn@doi [\aj]
  {10.1088/0004-6256/143/5/123}, \href
  {https://ui.adsabs.harvard.edu/abs/2012AJ....143..123M} {143, 123}

\bibitem[\protect\citeauthoryear{{Oda}, {Hino}, {Muto}, {Takahara}  \&
  {Sato}}{{Oda} et~al.}{1994}]{Oda1994}
{Oda} T.,  {Hino} M.,  {Muto} K.,  {Takahara} M.,   {Sato} K.,  1994, \mn@doi
  [Atomic Data and Nuclear Data Tables] {10.1006/adnd.1994.1007}, \href
  {http://adsabs.harvard.edu/abs/1994ADNDT..56..231O} {56, 231}

\bibitem[\protect\citeauthoryear{{Ogilvie}}{{Ogilvie}}{2014}]{2014ARA&A..52..171O}
{Ogilvie} G.~I.,  2014, \mn@doi [\araa] {10.1146/annurev-astro-081913-035941},
  \href {https://ui.adsabs.harvard.edu/abs/2014ARA%26A..52..171O} {52, 171}

\bibitem[\protect\citeauthoryear{Oliphant}{Oliphant}{2006}]{book}
Oliphant T.,  2006, Guide to NumPy

\bibitem[\protect\citeauthoryear{{Paxton}, {Bildsten}, {Dotter}, {Herwig},
  {Lesaffre}  \& {Timmes}}{{Paxton} et~al.}{2011}]{Paxton2011}
{Paxton} B.,  {Bildsten} L.,  {Dotter} A.,  {Herwig} F.,  {Lesaffre} P.,
  {Timmes} F.,  2011, \mn@doi [\apjs] {10.1088/0067-0049/192/1/3}, \href
  {http://adsabs.harvard.edu/abs/2011ApJS..192....3P} {192, 3}

\bibitem[\protect\citeauthoryear{{Paxton} et~al.,}{{Paxton}
  et~al.}{2013}]{Paxton2013}
{Paxton} B.,  et~al., 2013, \mn@doi [\apjs] {10.1088/0067-0049/208/1/4}, \href
  {http://adsabs.harvard.edu/abs/2013ApJS..208....4P} {208, 4}

\bibitem[\protect\citeauthoryear{{Paxton} et~al.,}{{Paxton}
  et~al.}{2015}]{Paxton2015}
{Paxton} B.,  et~al., 2015, \mn@doi [\apjs] {10.1088/0067-0049/220/1/15}, \href
  {http://adsabs.harvard.edu/abs/2015ApJS..220...15P} {220, 15}

\bibitem[\protect\citeauthoryear{{Paxton} et~al.,}{{Paxton}
  et~al.}{2018}]{Paxton2018}
{Paxton} B.,  et~al., 2018, \mn@doi [\apjs] {10.3847/1538-4365/aaa5a8}, \href
  {http://adsabs.harvard.edu/abs/2018ApJS..234...34P} {234, 34}

\bibitem[\protect\citeauthoryear{{Paxton} et~al.,}{{Paxton}
  et~al.}{2019}]{Paxton2019}
{Paxton} B.,  et~al., 2019, arXiv e-prints, \href
  {http://adsabs.harvard.edu/abs/2019arXiv190301426P} {}

\bibitem[\protect\citeauthoryear{{Pols}, {Tout}, {Eggleton}  \& {Han}}{{Pols}
  et~al.}{1995}]{Pols1995}
{Pols} O.~R.,  {Tout} C.~A.,  {Eggleton} P.~P.,   {Han} Z.,  1995, \mn@doi
  [\mnras] {10.1093/mnras/274.3.964}, \href
  {http://adsabs.harvard.edu/abs/1995MNRAS.274..964P} {274, 964}

\bibitem[\protect\citeauthoryear{{Potekhin} \& {Chabrier}}{{Potekhin} \&
  {Chabrier}}{2010}]{Potekhin2010}
{Potekhin} A.~Y.,  {Chabrier} G.,  2010, \mn@doi [Contributions to Plasma
  Physics] {10.1002/ctpp.201010017}, \href
  {http://adsabs.harvard.edu/abs/2010CoPP...50...82P} {50, 82}

\bibitem[\protect\citeauthoryear{{Rappaport}, {Deck}, {Levine}, {Borkovits},
  {Carter}, {El Mellah}, {Sanchis-Ojeda}  \& {Kalomeni}}{{Rappaport}
  et~al.}{2013}]{2013ApJ...768...33R}
{Rappaport} S.,  {Deck} K.,  {Levine} A.,  {Borkovits} T.,  {Carter} J.,  {El
  Mellah} I.,  {Sanchis-Ojeda} R.,   {Kalomeni} B.,  2013, \mn@doi [\apj]
  {10.1088/0004-637X/768/1/33}, \href
  {https://ui.adsabs.harvard.edu/abs/2013ApJ...768...33R} {768, 33}

\bibitem[\protect\citeauthoryear{{Remus}, {Mathis}  \& {Zahn}}{{Remus}
  et~al.}{2012}]{2012A&A...544A.132R}
{Remus} F.,  {Mathis} S.,   {Zahn} J.-P.,  2012, \mn@doi [\aap]
  {10.1051/0004-6361/201118160}, \href
  {https://ui.adsabs.harvard.edu/abs/2012A%26A...544A.132R} {544, A132}

\bibitem[\protect\citeauthoryear{{Rogers} \& {Nayfonov}}{{Rogers} \&
  {Nayfonov}}{2002}]{Rogers2002}
{Rogers} F.~J.,  {Nayfonov} A.,  2002, \mn@doi [\apj] {10.1086/341894}, \href
  {http://adsabs.harvard.edu/abs/2002ApJ...576.1064R} {576, 1064}

\bibitem[\protect\citeauthoryear{Rossum}{Rossum}{1995}]{Rossum:1995:PRM:869369}
Rossum G.,  1995, Technical report, Python Reference Manual.
Amsterdam, The Netherlands, The Netherlands

\bibitem[\protect\citeauthoryear{{Saumon}, {Chabrier}  \& {van Horn}}{{Saumon}
  et~al.}{1995}]{Saumon1995}
{Saumon} D.,  {Chabrier} G.,   {van Horn} H.~M.,  1995, \mn@doi [\apjs]
  {10.1086/192204}, \href {http://adsabs.harvard.edu/abs/1995ApJS...99..713S}
  {99, 713}

\bibitem[\protect\citeauthoryear{{Schou} et~al.,}{{Schou}
  et~al.}{1998}]{1998ApJ...505..390S}
{Schou} J.,  et~al., 1998, \mn@doi [\apj] {10.1086/306146}, \href
  {https://ui.adsabs.harvard.edu/abs/1998ApJ...505..390S} {505, 390}

\bibitem[\protect\citeauthoryear{{Southworth}, {Maxted}  \&
  {Smalley}}{{Southworth} et~al.}{2004}]{2004MNRAS.351.1277S}
{Southworth} J.,  {Maxted} P.~F.~L.,   {Smalley} B.,  2004, \mn@doi [\mnras]
  {10.1111/j.1365-2966.2004.07871.x}, \href
  {https://ui.adsabs.harvard.edu/abs/2004MNRAS.351.1277S} {351, 1277}

\bibitem[\protect\citeauthoryear{{Speagle}}{{Speagle}}{2019}]{2019arXiv190402180S}
{Speagle} J.~S.,  2019, arXiv e-prints, \href
  {https://ui.adsabs.harvard.edu/abs/2019arXiv190402180S} {}

\bibitem[\protect\citeauthoryear{{Spruit}}{{Spruit}}{2002}]{2002A&A...381..923S}
{Spruit} H.~C.,  2002, \mn@doi [\aap] {10.1051/0004-6361:20011465}, \href
  {https://ui.adsabs.harvard.edu/abs/2002A&A...381..923S} {381, 923}

\bibitem[\protect\citeauthoryear{{Timmes} \& {Swesty}}{{Timmes} \&
  {Swesty}}{2000}]{Timmes2000}
{Timmes} F.~X.,  {Swesty} F.~D.,  2000, \mn@doi [\apjs] {10.1086/313304}, \href
  {http://adsabs.harvard.edu/abs/2000ApJS..126..501T} {126, 501}

\bibitem[\protect\citeauthoryear{{Torres}, {Andersen}  \&
  {Gim{\'e}nez}}{{Torres} et~al.}{2010}]{2010A&ARv..18...67T}
{Torres} G.,  {Andersen} J.,   {Gim{\'e}nez} A.,  2010, \mn@doi [\aapr]
  {10.1007/s00159-009-0025-1}, \href
  {https://ui.adsabs.harvard.edu/abs/2010A%26ARv..18...67T} {18, 67}

\bibitem[\protect\citeauthoryear{{Zahn}}{{Zahn}}{1975}]{1975A&A....41..329Z}
{Zahn} J.~P.,  1975, \aap, \href
  {https://ui.adsabs.harvard.edu/abs/1975A&A....41..329Z} {41, 329}

\bibitem[\protect\citeauthoryear{{Zahn}}{{Zahn}}{1977}]{1977A&A....57..383Z}
{Zahn} J.-P.,  1977, \aap, \href
  {https://ui.adsabs.harvard.edu/abs/1977A%26A....57..383Z} {57, 383}

\makeatother
\end{thebibliography}

%%%%%%%%%%%%%%%%%%%%%%%%%%%%%%%%%%%%%%%%%%%%%%%%%%

%%%%%%%%%%%%%%%%% APPENDICES %%%%%%%%%%%%%%%%%%%%%

\appendix

\section{Torque Algebra}
\label{appen:torque}

We now compute the net torque on the star in the case where the lag angle $\alpha$ may vary with latitude.
To do so we cast equation~\eqref{eq:potential} in the form
\begin{align}
	\delta \Phi_{A\rightarrow B} \propto 3 (\hat{r}_{A\rightarrow B} \cdot \boldsymbol{r}_B)^2 - r_B^2
	\label{eq:tidal_phi}
\end{align}
\citep{1977A&A....57..383Z}, where $\hat{r}_{A\rightarrow B}$ is the unit vector pointing from the center of star $A$ to the center of star $B$.
The tidal acceleration is then
\begin{align}
	\boldsymbol{f} &= - \nabla \delta \Phi_{A\rightarrow B}\propto 6 (\hat{r}_{A\rightarrow B} \cdot \boldsymbol{r}_B) \hat{r}_{A\rightarrow B} - 2\boldsymbol{r}_B.
\end{align}
The specific torque on this fluid element relative to the center of mass of star $B$ is then
\begin{align}
	\boldsymbol{\tau} &= \boldsymbol{r}_B \times \boldsymbol{f}\propto 6 (\hat{r}_{A\rightarrow B} \cdot \boldsymbol{r}_B) \boldsymbol{r}_B\times\hat{r}_{A\rightarrow B}.
	\label{eq:tau}
\end{align}
We next make the coordinate transformation
\begin{align}
\boldsymbol{r}_B \rightarrow \boldsymbol{r}_B + \boldsymbol{\xi}(\boldsymbol{r}_B).
\end{align}
That is, we identify fluid elements by where they would have been in the absence of the tidal potential and explicitly account for the tidal bulge.
With this equation~\eqref{eq:tau} becomes
\begin{align}
	\boldsymbol{\tau} &\propto \left[\hat{r}_{A\rightarrow B} \cdot (\boldsymbol{r}_B + \boldsymbol{\xi})\right](\boldsymbol{r}_B + \boldsymbol{\xi})\times\hat{r}_{A\rightarrow B}.
	\label{eq:tau2}
\end{align}
where for notational compactness we have dropped the explicit dependence of $\boldsymbol{\xi}$ on $\boldsymbol{r}_B$.

Using equation~\eqref{eq:tau2} we find the net torque on star $B$ to be
\begin{align}
	\mathcal{T} = \int \boldsymbol{\tau} dm \propto \int \left[\hat{r}_{A\rightarrow B} \cdot (\boldsymbol{r}_B + \boldsymbol{\xi})\right](\boldsymbol{r}_B + \boldsymbol{\xi})\times\hat{r}_{A\rightarrow B} dm.
\end{align}
Note that if the tidal displacement were zero the torque integrated over the whole star would vanish, so we may neglect the first term and expand the remaining terms to linear order in the displacement to find
\begin{align}
	\mathcal{T} &\propto \hat{r}_{A\rightarrow B} \cdot \int \left(\boldsymbol{r}_B \otimes \boldsymbol{\xi} + \boldsymbol{\xi}\otimes \boldsymbol{r}_B\right) dm\times\hat{r}_{A\rightarrow B},
	\label{eq:T0}
\end{align}
where $\otimes$ denotes an outer product.

When $\alpha = 0$ the system is symmetric with respect to reflection about $\hat{r}_{A\rightarrow B}$.
It follows that the displacement $\boldsymbol{\xi}_{\rm eq}$ is antisymmetric with respect to reflection about the same.
Because the torque on a fluid element is linear in $\boldsymbol{\xi}$, this means that when $\alpha=0$ the net torque vanishes.
Hence
\begin{align}
	\mathcal{T} &= \int \boldsymbol{\tau} - \boldsymbol{\tau}_{\alpha=0} dm\approx \int \alpha \left.\frac{d\boldsymbol{\tau}}{d\alpha}\right|_{\alpha=0} dm,
\end{align}
where we have expanded $\mathcal{T}$ to leading order in $\alpha$.

To compute $d\boldsymbol{\tau}/d\alpha|_{\alpha=0}$ we note that in equation~\eqref{eq:T0} only $\boldsymbol{\xi}$ depends on $\alpha$.
Moreover
\begin{align}
	\left.\frac{\partial\boldsymbol{\xi}}{\partial\alpha}(\boldsymbol{r}_B,t)\right|_{\alpha=0} = \frac{1}{\Omega-\omega}\frac{d\boldsymbol{\xi}}{dt} = \frac{1}{\Omega-\omega}\frac{d\boldsymbol{r}_B}{dt} \cdot \nabla \boldsymbol{\xi}_{\rm eq}.
\end{align}
So
\begin{align}
	\mathcal{T} &\propto \hat{r}_{A\rightarrow B} \cdot \int \left(\boldsymbol{r}_B \otimes \left.\frac{\partial\boldsymbol{\xi}}{\partial \alpha}\right|_{\alpha=0} + \left.\frac{\partial\boldsymbol{\xi}}{\partial \alpha}\right|_{\alpha=0}\otimes \boldsymbol{r}_B\right) dm\times\hat{r}_{A\rightarrow B}\\
				&= \hat{r}_{A\rightarrow B} \cdot \int \left(\frac{\alpha}{\Omega-\omega} \boldsymbol{r}_B \otimes \frac{d\boldsymbol{\xi}_{\rm eq}}{dt} + \mathrm{transpose}\right)dm\times\hat{r}_{A\rightarrow B}.
\end{align}

In equilibrium (i.e. with $\hat{r}_{A\rightarrow B}$, $\Omega$ and $\omega$ time-independent), $\mathcal{T}$ does not depend on time.
This is because with equations~\eqref{eq:r_rot} and~\eqref{eq:xi} the properties of the fluid are time-independent in the Eulerian sense, even as fluid elements travel from place to place.
Thus if we expand $dm = \rho d^3\boldsymbol{r}$ and integrate over space instead of mass we must obtain a constant.

\begin{figure}
\centering
\includegraphics[width=0.4\textwidth]{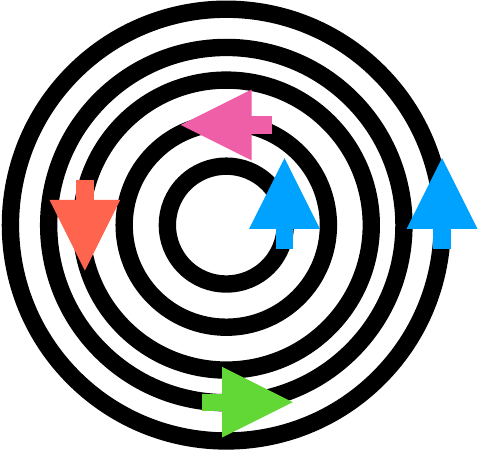}
\caption{Rings of material which rotate into the same Eulerian positions are shown. Note that these may have different rotation rates from one another, but we require that that rate must be uniform within each ring.}
\label{fig:schema3}
\end{figure}

In fact an even stronger statement holds.
Consider the ring of fluid elements which at various points in time each occupy the same position owing to the star's rotation.
Several of these are shown schematically in Figure~\ref{fig:schema3}.
Each ring is time-invariant in the Eulerian sense, so the net torque on each must be time-independent too.
Because the net torque on such a ring is constant it is unchanged if we average it over a time $2\pi (\Omega-\omega)^{-1}$.
Hence
\begin{align}
	\mathcal{T} &\propto \hat{r}_{A\rightarrow B} \cdot \int \alpha\int_0^{\frac{2\pi}{\Omega-\omega}}\left( \boldsymbol{r}_B \otimes \frac{d\boldsymbol{\xi}_{\rm eq}}{dt} + \mathrm{transpose}\right)\,dt dm\times\hat{r}_{A\rightarrow B}.
\end{align}
The time integration may be done by parts for each fluid element, so that
\begin{align}
\int_0^{\frac{2\pi}{\Omega-\omega}} \boldsymbol{r}_B \otimes \frac{d\boldsymbol{\xi}_{\rm eq}}{dt} dt &= \left.\boldsymbol{r}_B\otimes  \boldsymbol{\xi}_{\rm eq}\right|_0^{\frac{2\pi}{\Omega-\omega}}-\int_0^{\frac{2\pi}{\Omega-\omega}} \frac{d\boldsymbol{r}_B}{dt} \otimes \boldsymbol{\xi}_{\rm eq} dt.
\end{align}
Noting that for any fluid element $\boldsymbol{r}_B$ and $\boldsymbol{\xi}$ are periodic in time with the same period we see that the first term vanishes, so
\begin{align}
\int_0^{\frac{2\pi}{\Omega-\omega}} \boldsymbol{r}_B \otimes \frac{d\boldsymbol{\xi}_{\rm eq}}{dt} dt &= -\int_0^{\frac{2\pi}{\Omega-\omega}} \frac{d\boldsymbol{r}_B}{dt} \otimes \boldsymbol{\xi}_{\rm eq} dt.
\end{align}
Thus
\begin{align}
	\mathcal{T} &\propto \hat{r}_{A\rightarrow B} \cdot \int \left(\alpha\int_0^{\frac{2\pi}{\Omega-\omega}} \frac{d\boldsymbol{r}_B}{dt} \otimes \boldsymbol{\xi}_{\rm eq} + \mathrm{transpose}\right)dt dm\times\hat{r}_{A\rightarrow B}.
\end{align}
Because the integrand is actually time-independent when summed over one of the rings in Figure~\ref{fig:schema3}, and because this summation is done by the integral over $dm$, we may drop the time average to obtain
\begin{align}
	\mathcal{T} &\propto \hat{r}_{A\rightarrow B} \cdot \int \left(\frac{\alpha}{\Omega-\omega}\frac{d\boldsymbol{r}_B}{dt} \otimes \boldsymbol{\xi}_{\rm eq} + \mathrm{transpose}\right) dm\times\hat{r}_{A\rightarrow B}.
	\label{eq:T1}
\end{align}
Recalling equation~\eqref{eq:r_rot} we have
\begin{align}
	\frac{d\boldsymbol{r}_B}{dt} = (\Omega-\omega)\hat{z} \times \boldsymbol{r}_B,
\end{align}
so equation~\eqref{eq:T1} becomes
\begin{align}
	\mathcal{T} &\propto \hat{r}_{A\rightarrow B} \cdot \int \left(\alpha \hat{z} \times \boldsymbol{r}_B \otimes \boldsymbol{\xi}_{\rm eq} + \mathrm{transpose}\right) dm\times\hat{r}_{A\rightarrow B}.
\end{align}

Because we have assumed that the orbital and spin axes are aligned the torque can only be along $\hat{z}$.
It follows that
\begin{align}
	\mathcal{T} &\propto \hat{z} \left[\hat{r}_{A\rightarrow B} \cdot\int \left( \alpha \hat{z} \times \boldsymbol{r}_B \otimes \boldsymbol{\xi}_{\rm eq} + \mathrm{transpose}\right) dm\times\hat{r}_{A\rightarrow B}\right]\cdot\hat{z}.
\end{align}
The unit vectors $\hat{r}_{A\rightarrow B}$ and $\hat{z}$ are perpendicular, so the only contribution to this torque comes from components of the integral which are along $\hat{r}_{A\rightarrow B} \times \hat{z}$, which we shall call $\hat{y}$ for convenience.
Thus
\begin{align}
	\mathcal{T} &\propto \hat{z} \left[\hat{r}_{A\rightarrow B} \cdot \int \left(\alpha \hat{z} \times \boldsymbol{r}_B \otimes \boldsymbol{\xi}_{\rm eq} + \mathrm{transpose}\right) dm\cdot \hat{y}\right]\\
	&\propto \hat{z} \left[\hat{r}_{B\rightarrow A} \cdot \int \left(\alpha \hat{z} \times \boldsymbol{r}_B \otimes \boldsymbol{\xi}_{\rm eq} + \mathrm{transpose}\right) dm\cdot \hat{y}\right],
	\label{eq:T2}
\end{align}
where $\hat{r}_{B\rightarrow A} = -\hat{r}_{A\rightarrow B}$ is the unit vector pointing from star $B$ to star $A$.

\begin{figure}
\centering
\includegraphics[width=0.4\textwidth]{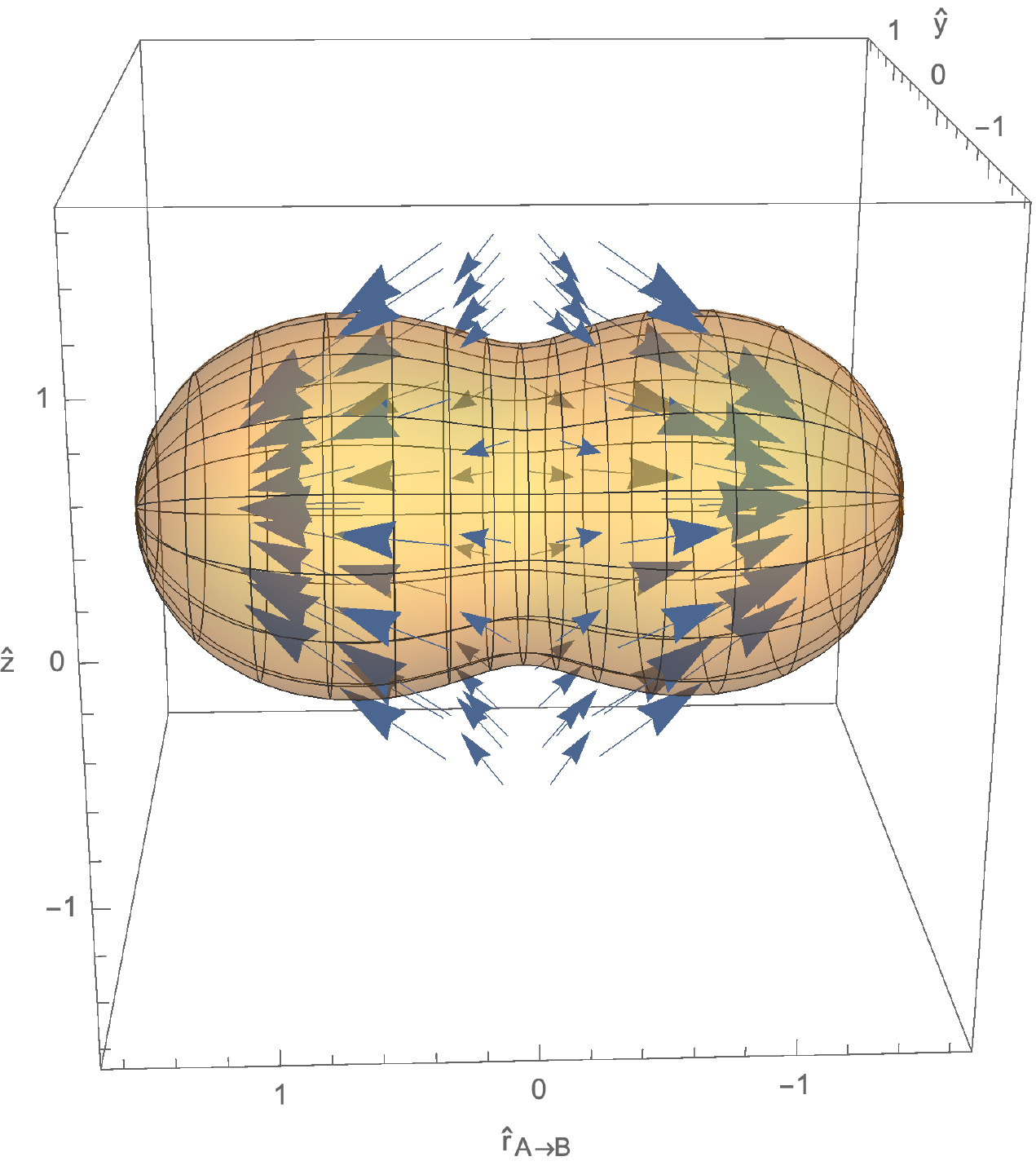}
\caption{The surface of the tidally-distorted star and the displacement field are plotted schematically, with the amplitude of the distortion exaggerated for clarity. The horizontal axis is $\hat{r}_{A\rightarrow B}$,  the vertical is $\hat{z}$ (the orbital axis) and the one running into the page is $\hat{y}$.}
\label{fig:schema4}
\end{figure}

To proceed we must determine the equilibrium tidal displacement.
The radial component is given approximately by
\begin{align}
	\xi_{r,{\rm eq}} \approx \frac{\delta \Phi_{A\rightarrow B}}{g_B}
	\label{eq:xir}
\end{align}
\citep{2012A&A...544A.132R},
where $g_B$ is the inward-pointing component of the unperturbed gravitational field of star $B$.
The remaining components are fixed by the condition that
\begin{align}
	\nabla\cdot\boldsymbol{\xi}_{\rm eq} = 0
	\label{eq:xi0}
\end{align}
\citep{2012A&A...544A.132R}.
To solve this equation we define $\psi$ to be the angle between $\boldsymbol{r}_B$ and $\hat{r}_{B\rightarrow A}$, and $\phi$ to be the rotation angle about $\hat{r}_{A\rightarrow B}$ relative to $\hat{z}$, as shown in Figure~\ref{fig:schema5}.
With that, 
\begin{align}
\delta \Phi_{A\rightarrow B} \propto r_B^2 Y_{2,0}(\psi,\phi) \propto r_B^2 (3\cos^2\psi - 1),
\end{align}
where $Y_{lm}$ are the spherical harmonics.
We further note that by symmetry there is no displacement about the $\hat{r}_{A\rightarrow B}$ axis.
So the solution to equation~\eqref{eq:xi0} with equation~\eqref{eq:xir} yields
\begin{align}
	\boldsymbol{\xi}_{\rm eq} &\propto \hat{r}_B Y_{2,0} \frac{r_B^2}{g} + \frac{r_B}{6}\left[\frac{2 r_B}{g} + \frac{d}{d r_B}\left(\frac{r_B^2}{g}\right) \nabla Y_{2,0}\right].
\end{align}
Except near the core of the star it is a good approximation to say that $g \propto r^{-2}$.
With this we find
\begin{align}
	\boldsymbol{\xi}_{\rm eq} &\propto \frac{r_B^2}{g}\left[\hat{r}_B (3\cos^2\psi - 1) - 3 \sin(2\psi)\hat{\psi}\right].
\end{align}
This displacement field is plotted schematically in Figure~\ref{fig:schema4}, showing that this tide consists primarly of material flowing along the $\hat{r}_{A\rightarrow B}$ axis away from the center of mass of star $B$ in both directions.

To compute the torque then we note that
\begin{align}
	\hat{r}_B \cdot \hat{r}_{B\rightarrow A} &= \cos\psi\\
	\hat{r}_B \cdot \hat{y} &= \sin\psi \sin\phi\\
	\hat{r}_B \cdot \hat{z} &= \sin\psi \cos\phi\\
	(\hat{z} \times \hat{r}_B) \cdot \hat{r}_{B\rightarrow A} &= \sin\psi \sin\phi\\
	(\hat{z} \times \hat{r}_B) \cdot \hat{y} &= -\cos\psi\\
	(\hat{z} \times \hat{r}_B) \cdot \hat{z} &= 0\\
	\hat{\psi} \cdot \hat{r}_{B\rightarrow A} &= -\sin\psi\\
	\hat{\psi} \cdot \hat{y} &= \cos\psi \sin\phi\\
	\hat{\psi} \cdot \hat{z} &= \cos\psi\cos\phi,
\end{align}
so
\begin{align}
	\boldsymbol{\xi}_{\rm eq} \cdot \hat{r}_{B\rightarrow A} &\propto \frac{r_B^2}{g}\left[(3\cos^2\psi - 1)\cos\psi + 3 \sin(2\psi)\sin\psi\right]\\
	\boldsymbol{\xi}_{\rm eq} \cdot \hat{y} &\propto \frac{r_B^2}{g}\left[(3\cos^2\psi - 1)\sin\psi \sin\phi - 3 \sin(2\psi)\cos\psi \sin\phi\right]\\
	\boldsymbol{\xi}_{\rm eq} \cdot \hat{z} &\propto \frac{r_B^2}{g}\left[(3\cos^2\psi - 1)\sin\psi \cos\phi - 3 \sin(2\psi)\cos\psi \cos\phi\right].
\end{align}
Inserting this into equation~\eqref{eq:T2} we find
\begin{align}
	\mathcal{T} &\propto \hat{z} \left[\hat{r}_{B\rightarrow A} \cdot \int \alpha \hat{z} \times \boldsymbol{r}_B \otimes \boldsymbol{\xi}_{\rm eq}dm\right] \cdot \hat{y}\nonumber\\
	&+\hat{z} \left[\hat{r}_{B\rightarrow A} \cdot \int \alpha \boldsymbol{\xi}_{\rm eq} \otimes \hat{z} \times \boldsymbol{r}_Bdm\right] \cdot \hat{y}\\
&\propto \hat{z} \int \alpha \sin\psi \sin\phi \frac{r_B^3}{g}[(3\cos^2\psi - 1)\sin\psi \sin\phi\nonumber\\
&\,\,\,\,\, - 3 \sin(2\psi)\cos\psi \sin\phi]dm\nonumber\\
&\,\,\,\,\,-\hat{z}\int \alpha \frac{r_B^3}{g}\left[(3\cos^2\psi - 1)\cos\psi + 3 \sin(2\psi)\sin\psi\right]\cos\psi dm\\	
&\propto\hat{z}\int \alpha \frac{r_B^3}{g}\left[\cos\psi\left(\cos\psi - 3\cos^3\psi - 3\cos(2\psi)\sin\psi\right)\right.\nonumber\\
&\,\,\,\,\,\left.-\frac{1}{4}\sin^2\phi \sin\psi \big(6\cos\psi + 6\cos(3\psi) + \sin\psi - 3 \sin(3\psi)\big)\right]dm.
\end{align}
Finally using 
\begin{align}
dm = \rho d^3\boldsymbol{r}_B = \rho r_B^2 dr_B \sin\psi d\psi d\phi
\end{align}
we obtain
\begin{align}
\mathcal{T}&\propto\hat{z}\int_0^{R_B}dr_B \int_0^{\pi}d\psi \sin\psi \int_0^{2\pi} d\phi \rho \alpha \frac{r_B^5}{g}\nonumber\\
&\,\,\,\,\,\times\left[\cos\psi\left(\cos\psi - 3\cos^3\psi - 3\cos(2\psi)\sin\psi\right)\right.\nonumber\\
&\,\,\,\,\,\left.-\frac{1}{4}\sin^2\phi \sin\psi \big(6\cos\psi + 6\cos(3\psi) + \sin\psi - 3 \sin(3\psi)\big)\right],
\label{eq:T3}
\end{align}
where $R_B$ is the outer radius of star $B$ in the absence of tidal perturbations and $\rho$ is the unperturbed density.

\section{Microphysics}
\label{appen:mesa}

The \code{MESA} equation of state (EOS) is a blend of the OPAL~\citep{Rogers2002}, SCVH~\citep{Saumon1995}, PTEH~\citep{Pols1995}, HELM~\citep{Timmes2000}, and PC~\citep{Potekhin2010} EOSes.
Radiative opacities are primarily from OPAL~\citep{Iglesias1993,
Iglesias1996}, with low-temperature data from~\citet{Ferguson2005}
and the high-temperature, Compton-scattering dominated regime by~\citet{Buchler1976}.
Electron conduction opacities are from~\citet{Cassisi2007}.
Nuclear reaction rates are from JINA REACLIB~\citep{Cyburt2010} plus additional
tabulated weak reaction rates~\citep{Fuller1985, Oda1994, Langanke2000}.
Screening is included via the prescription of~\citet{Chugunov2007}.
Thermal neutrino loss rates are from~\citet{Itoh1996}.

\section{Inclusive Fits}
\label{appen:yes_outliers}

Here we provide the results of the fits described in Section~\ref{sec:infer} which do not exclude outliers.

\begin{figure}
	\includegraphics[width=0.5\textwidth]{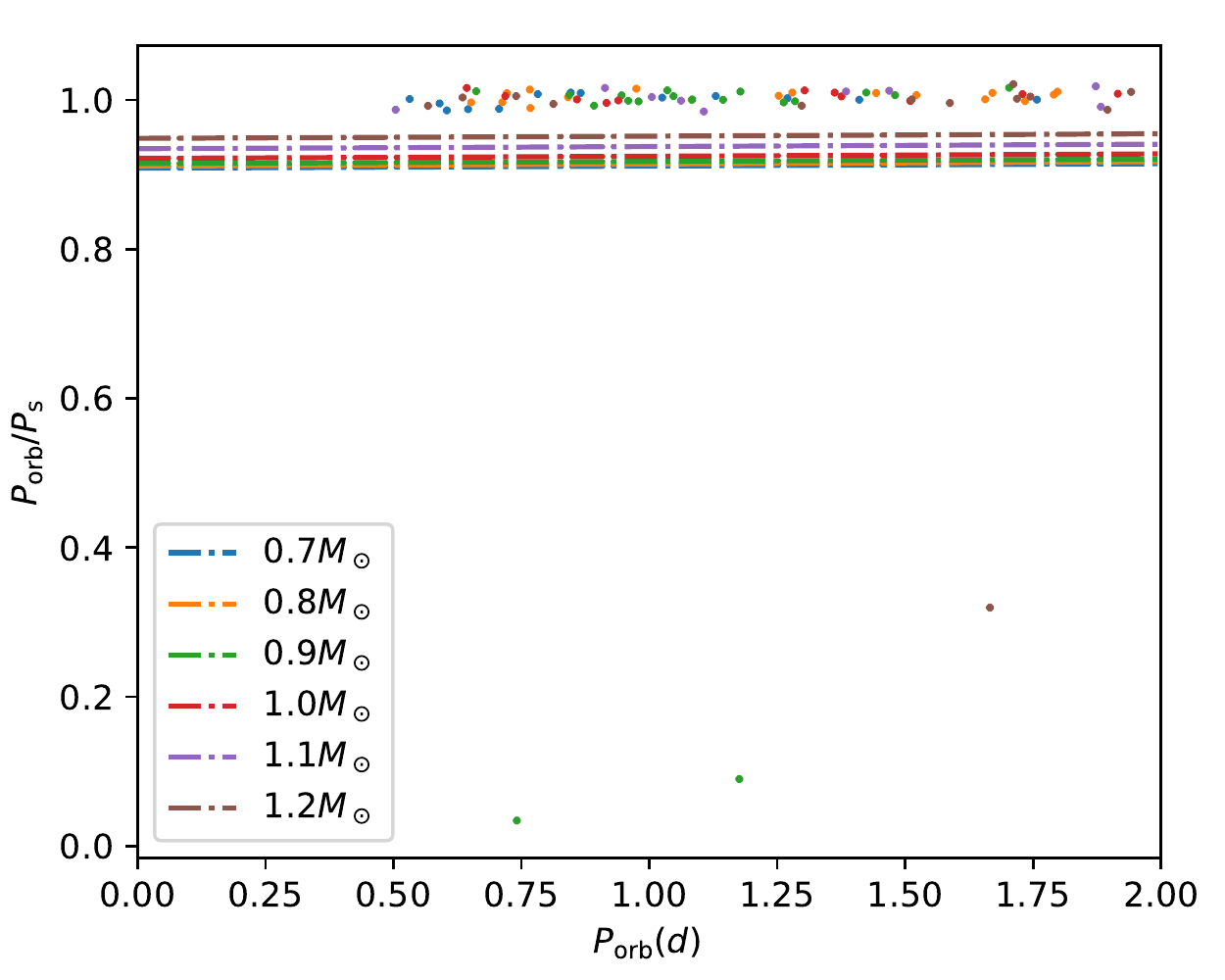}
	\caption{The ratio of orbital period to rotation period is shown as a function of the orbital period various masses, indicated by color. These results are from the inference including systems with orbital periods less than $2\,{\rm d}$. The dashed lines indicate the model predictions at the median parameter values for the stellar masses of the same color.}
	\label{fig:02d_with_outliers}
\end{figure}

In each case there are some points which are significantly offset from the model predictions (see e.g. Figure~\ref{fig:02d_with_outliers}).
This suggests that these fits include some systems which are not well-captured by our model.
It also explains the large effect the inclusion of these outliers has on the fit parameters (see Table~\ref{tab:res}).

%-----------

\begin{figure}
	\includegraphics[width=0.5\textwidth]{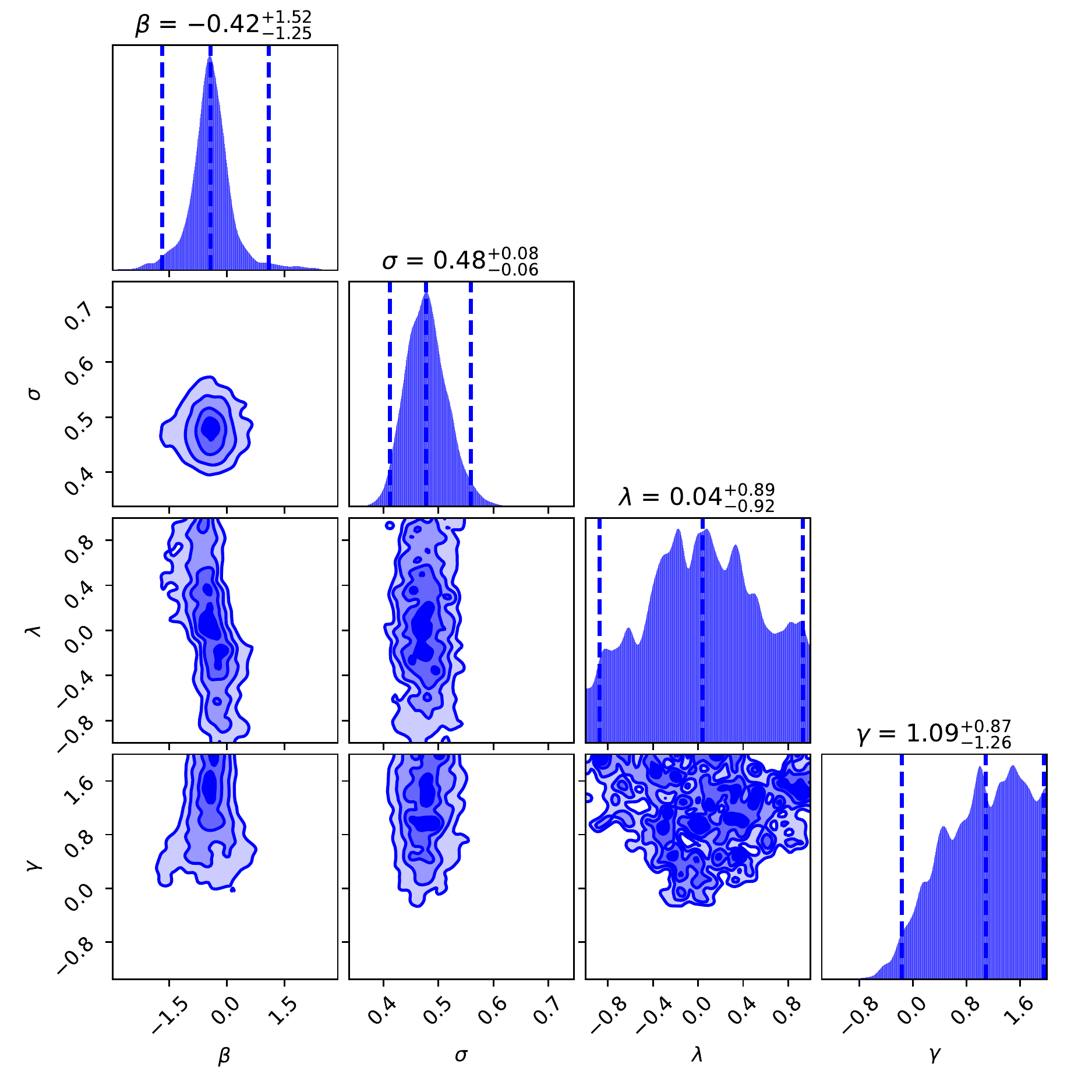}
	\caption{The marginalized posterior distribution over our model parameters is shown for the inference including systems with orbital periods less than $2\,{\rm d}$.}
\end{figure}

%-----------

\begin{figure}
	\includegraphics[width=0.5\textwidth]{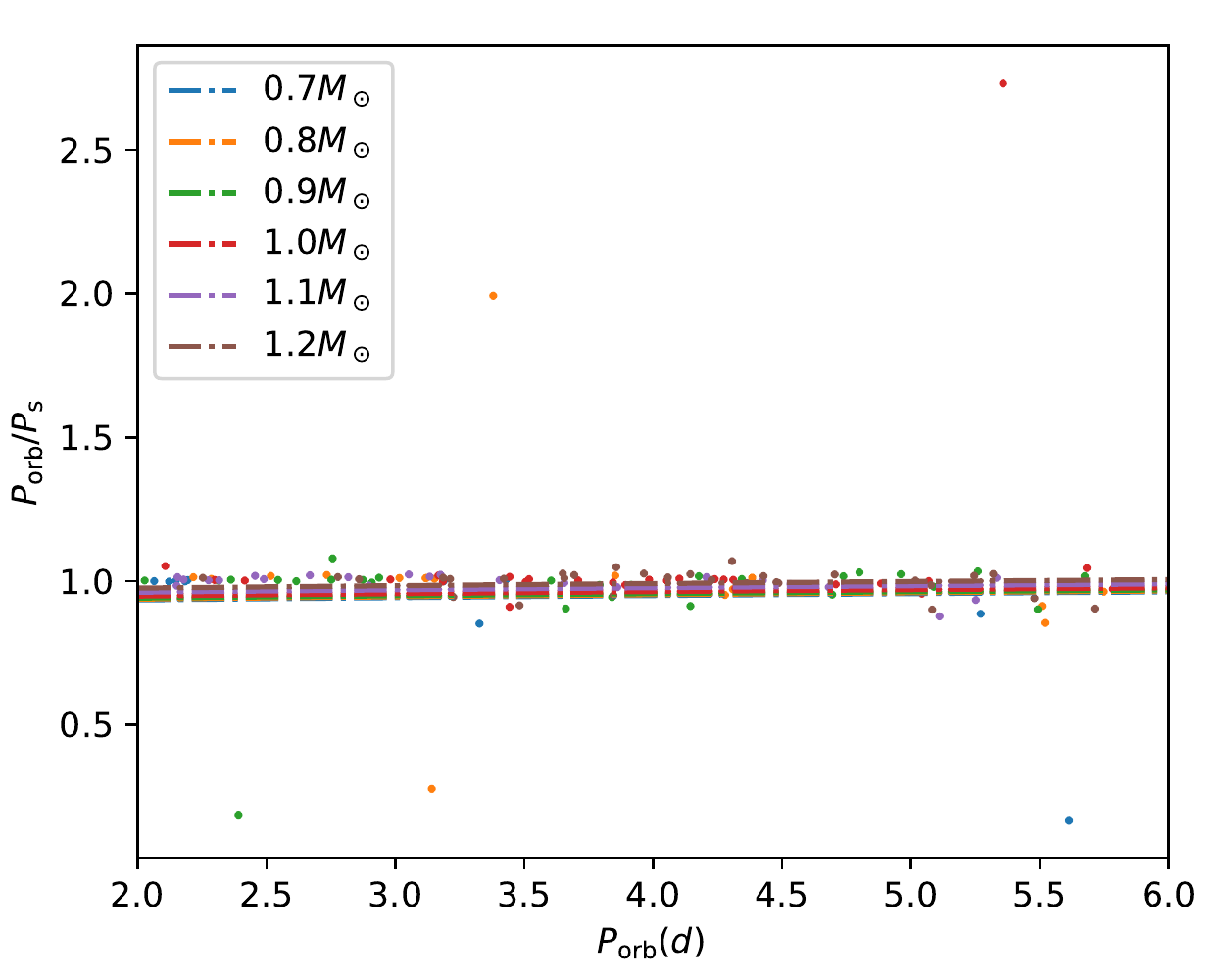}
	\caption{The ratio of orbital period to rotation period is shown as a function of the orbital period various masses, indicated by color. These results are from the inference including systems with orbital periods between $2\,{\rm d}$ and $6\,{\rm d}$. The dashed lines indicate the model predictions at the median parameter values for the stellar masses of the same color.}
\end{figure}

\begin{figure}
	\includegraphics[width=0.5\textwidth]{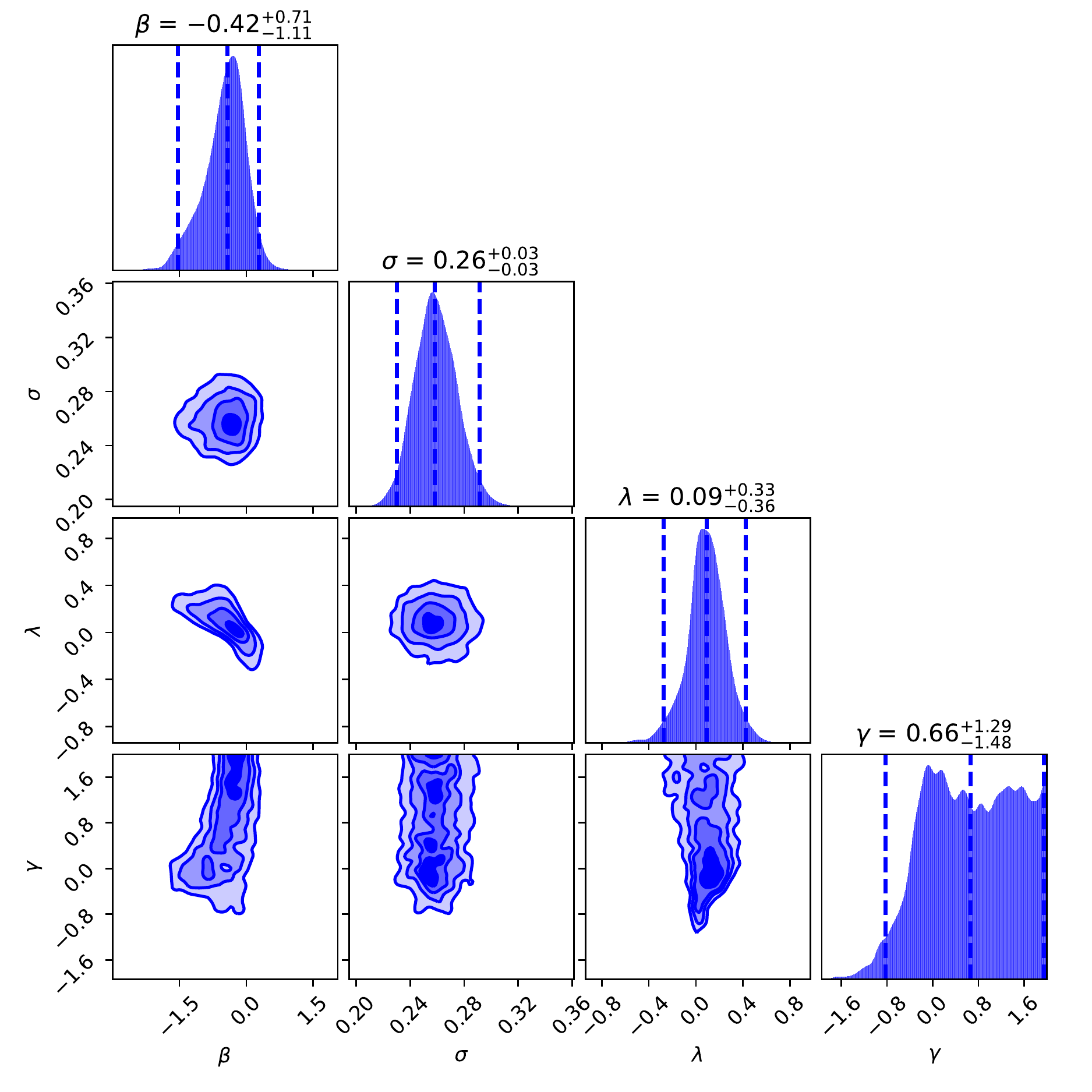}
	\caption{The marginalized posterior distribution over our model parameters is shown for the inference including systems with orbital periods between $2\,{\rm d}$ and $6\,{\rm d}$.}
\end{figure}

%-----------

\begin{figure}
	\includegraphics[width=0.5\textwidth]{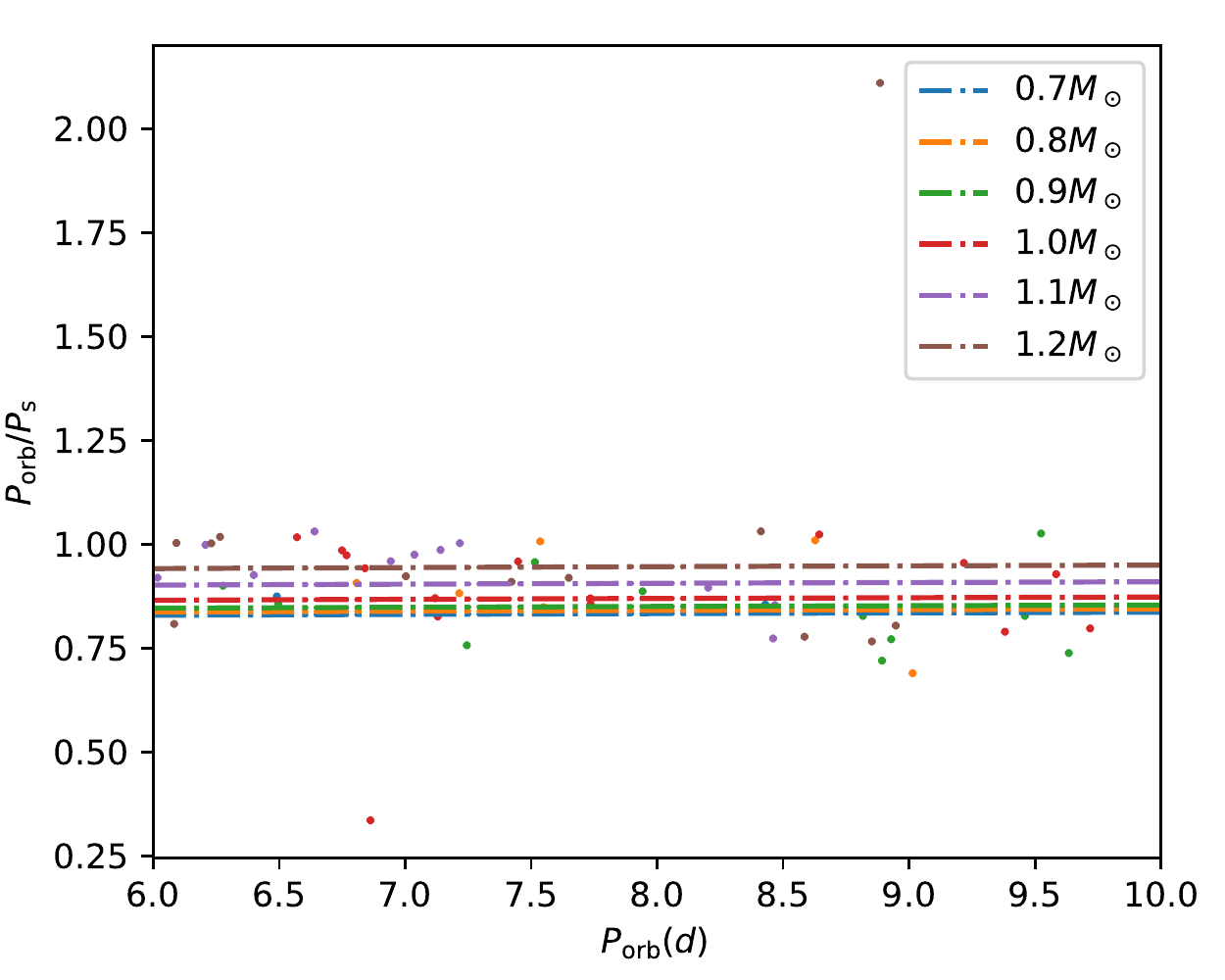}
	\caption{The ratio of orbital period to rotation period is shown as a function of the orbital period various masses, indicated by color. These results are from the inference including systems with orbital periods between $6\,{\rm d}$ and $10\,{\rm d}$. The dashed lines indicate the model predictions at the median parameter values for the stellar masses of the same color.}
\end{figure}

\begin{figure}
	\includegraphics[width=0.5\textwidth]{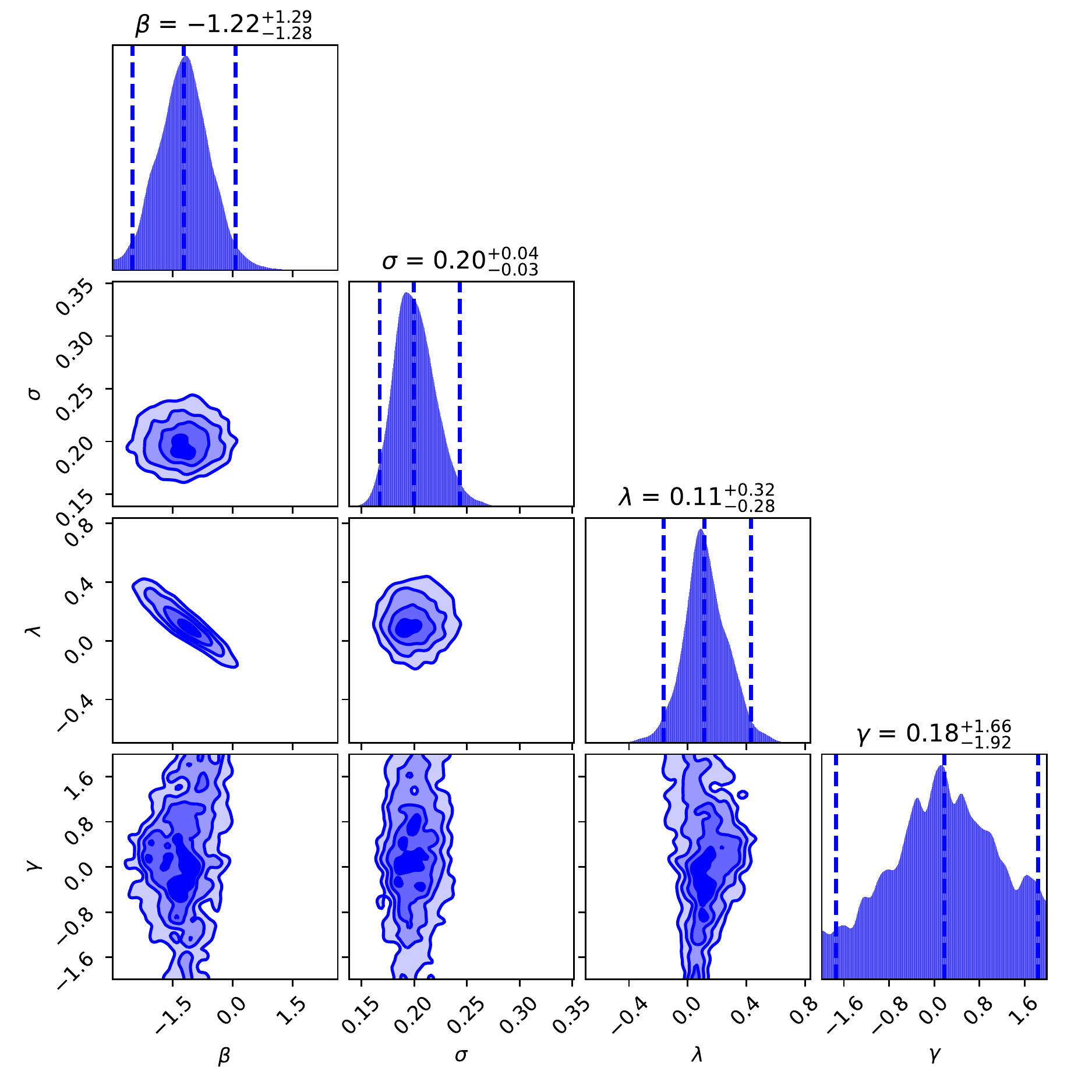}
	\caption{The marginalized posterior distribution over our model parameters is shown for the inference including systems with orbital periods between $6\,{\rm d}$ and $10\,{\rm d}$.}
\end{figure}

%-----------

\begin{figure}
	\includegraphics[width=0.5\textwidth]{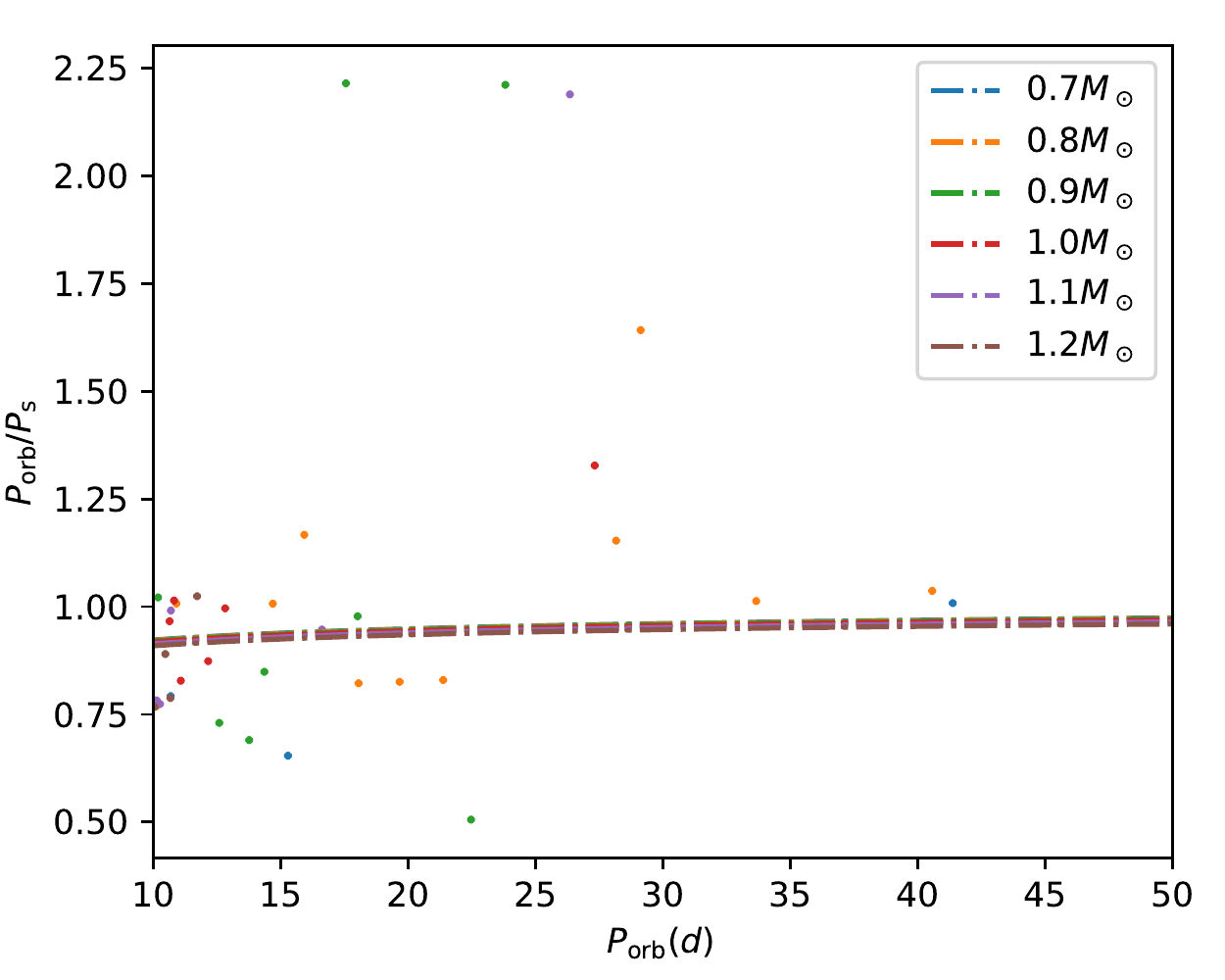}
	\caption{The ratio of orbital period to rotation period is shown as a function of the orbital period various masses, indicated by color. These results are from the inference including systems with orbital periods between $10\,{\rm d}$ and $50\,{\rm d}$. The dashed lines indicate the model predictions at the median parameter values for the stellar masses of the same color.}
\end{figure}

\begin{figure}
	\includegraphics[width=0.5\textwidth]{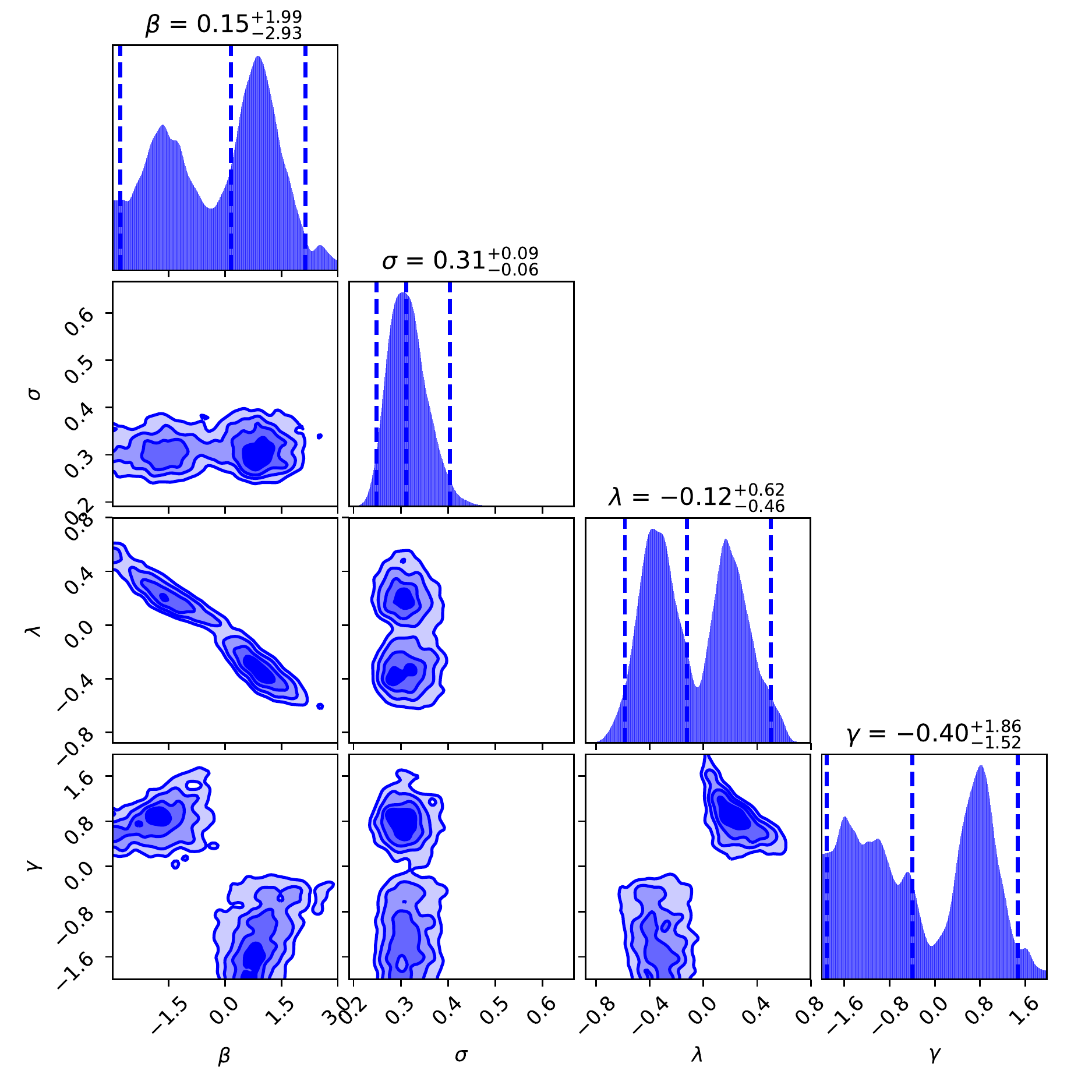}
	\caption{The marginalized posterior distribution over our model parameters is shown for the inference including systems with orbital periods between $10\,{\rm d}$ and $50\,{\rm d}$.}
\end{figure}

\newcommand{\po}{P_{\rm orb} > 2\,{\rm d}}
\section{$\protect\po$}
\label{appen:full}

Here we provide the results of the fit described in Section~\ref{sec:infer} for the period windows with $P_{\rm orb} > 2\,{\rm d}$ excluding outliers.

\begin{figure}
	\includegraphics[width=0.5\textwidth]{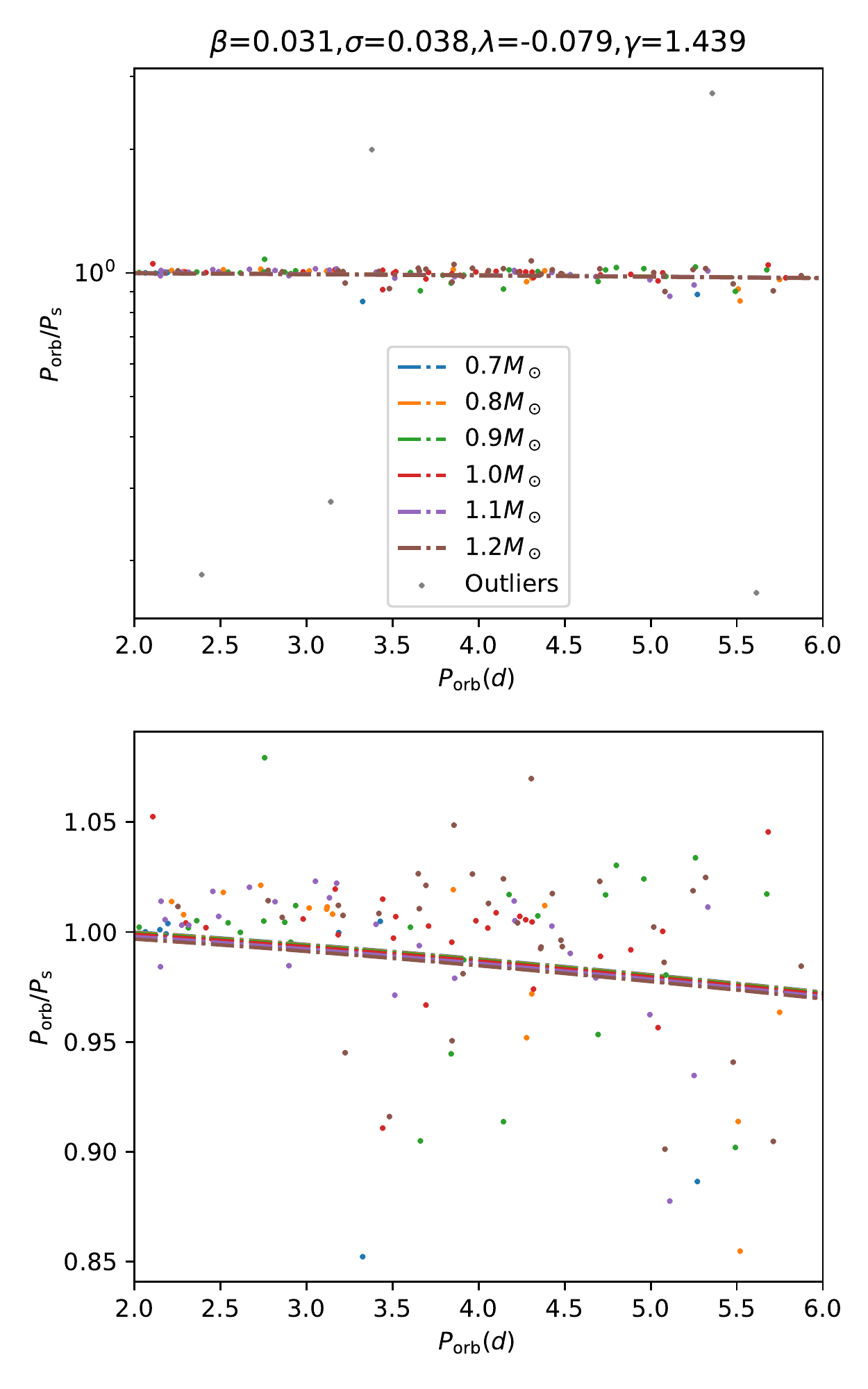}
	\caption{The ratio of orbital period to rotation period is shown as a function of the orbital period and stellar mass, indicated by color. Outliers were identified by the inference procedure as systems with a likelihood at the posterior median parameter values below $10^{-6}$, and are shown as grey circles. The upper panel shows all objects while the lower excludes outliers. These results are from the inference including systems with orbital periods between $2\,{\rm d}$ and $6\,{\rm d}$. The dashed lines indicate the model predictions at the median parameter values for the stellar masses of the same color.}
	\label{fig:26d}
\end{figure}

\begin{figure}
	\includegraphics[width=0.5\textwidth]{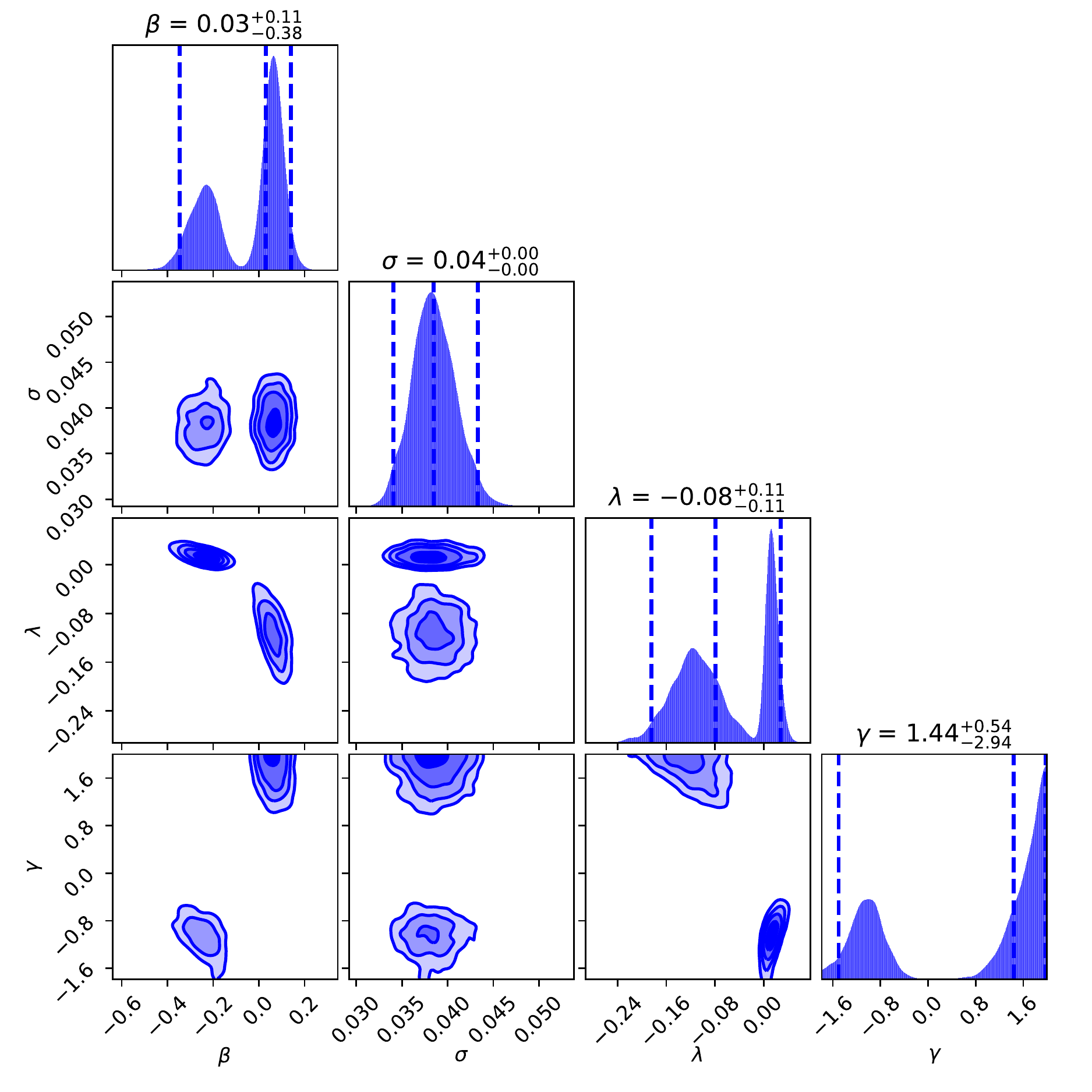}
	\caption{The marginalized posterior distribution over our model parameters is shown for the inference including systems with orbital periods between $2\,{\rm d}$ and $6\,{\rm d}$.}
	\label{fig:26d_post}
\end{figure}

\begin{figure}
	\includegraphics[width=0.5\textwidth]{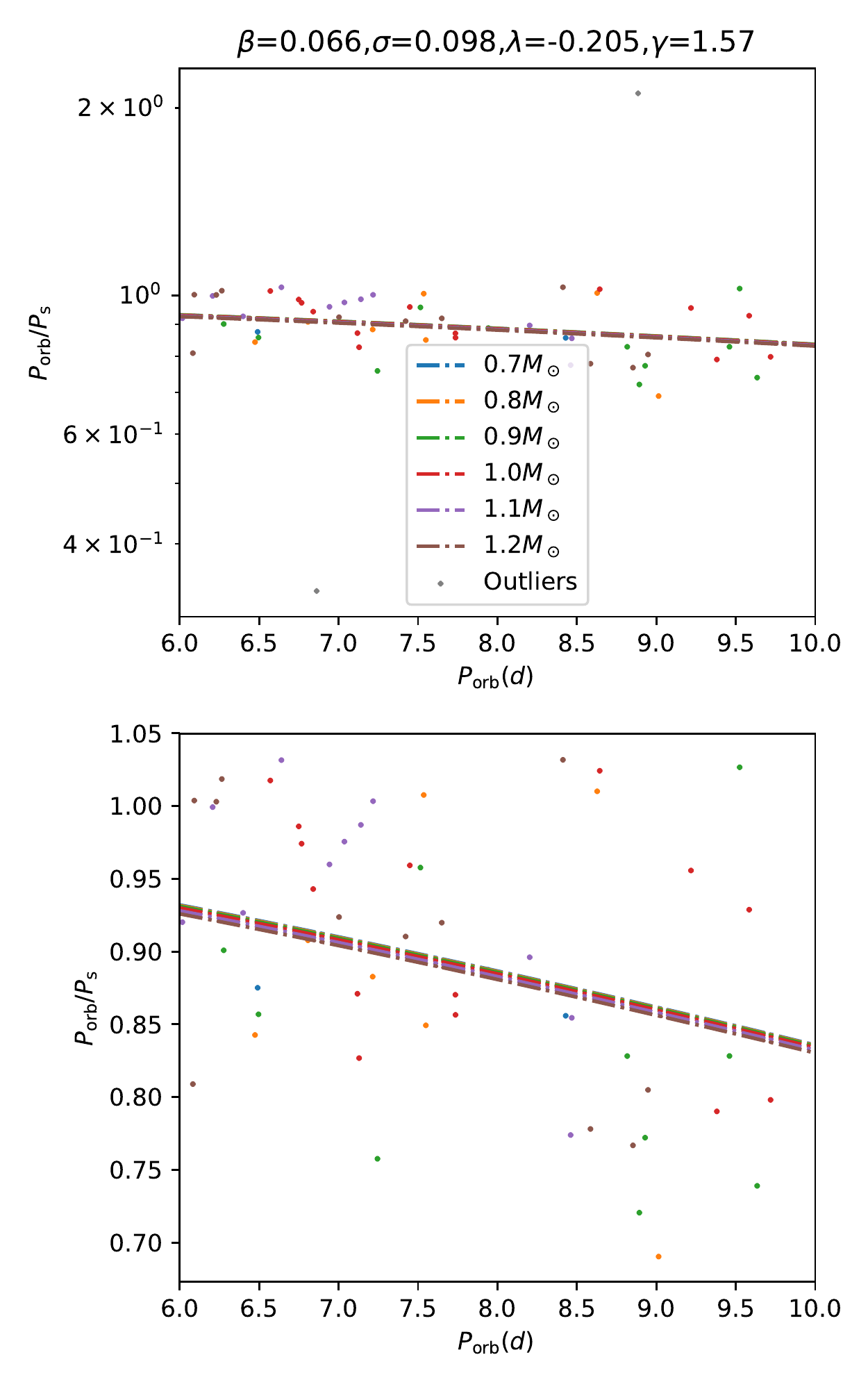}
	\caption{The ratio of orbital period to rotation period is shown as a function of the orbital period and stellar mass, indicated by color. Outliers were identified by the inference procedure as systems with a likelihood at the posterior median parameter values below $10^{-6}$, and are shown as grey circles. The upper panel shows all objects while the lower excludes outliers. These results are from the inference including systems with orbital periods between $6\,{\rm d}$ and $10\,{\rm d}$. The dashed lines indicate the model predictions at the median parameter values for the stellar masses of the same color.}
	\label{fig:610d}
\end{figure}

\begin{figure}
	\includegraphics[width=0.5\textwidth]{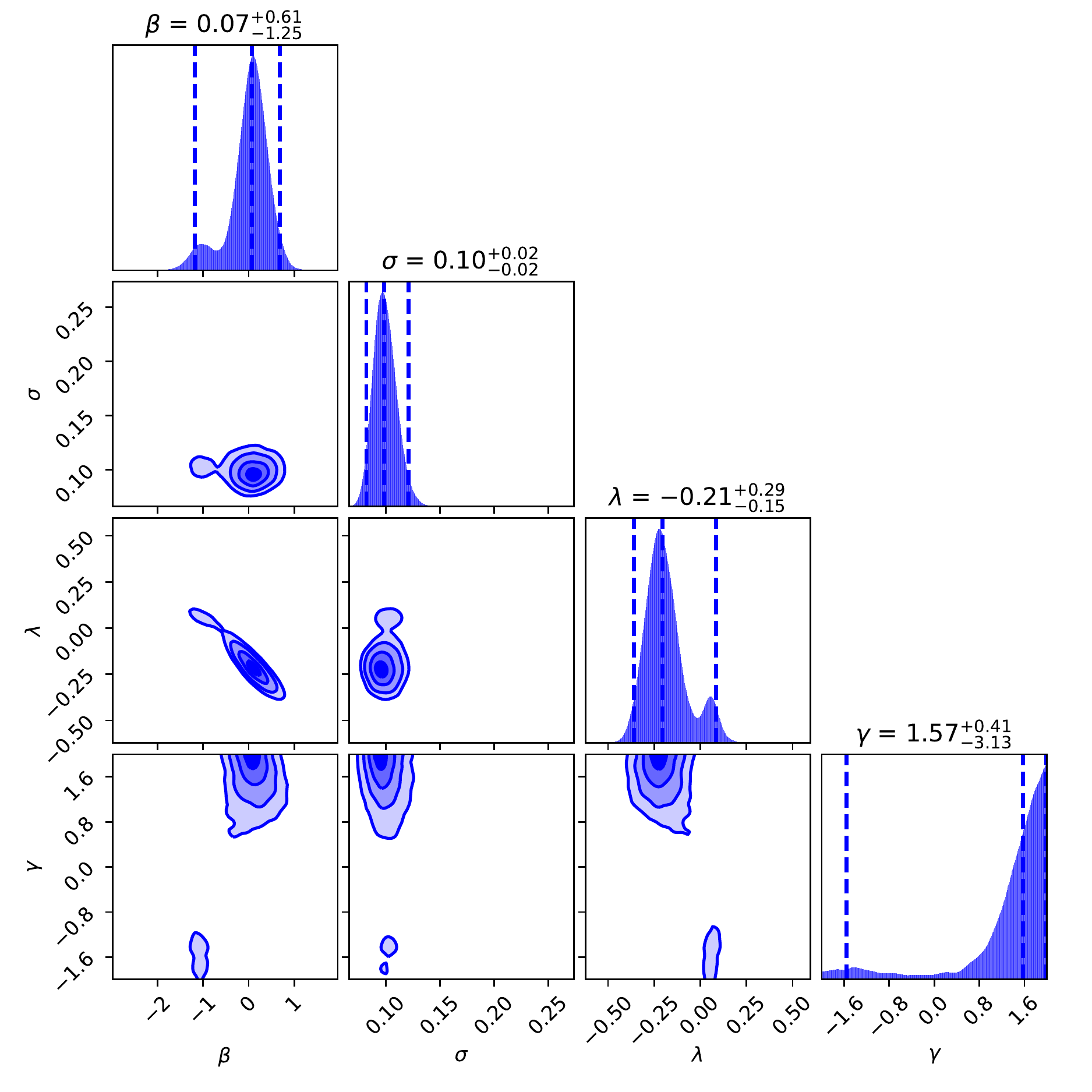}
	\caption{The marginalized posterior distribution over our model parameters is shown for the inference including systems with orbital periods between $6\,{\rm d}$ and $10\,{\rm d}$.}
	\label{fig:610d_post}
\end{figure}

\begin{figure}
	\includegraphics[width=0.5\textwidth]{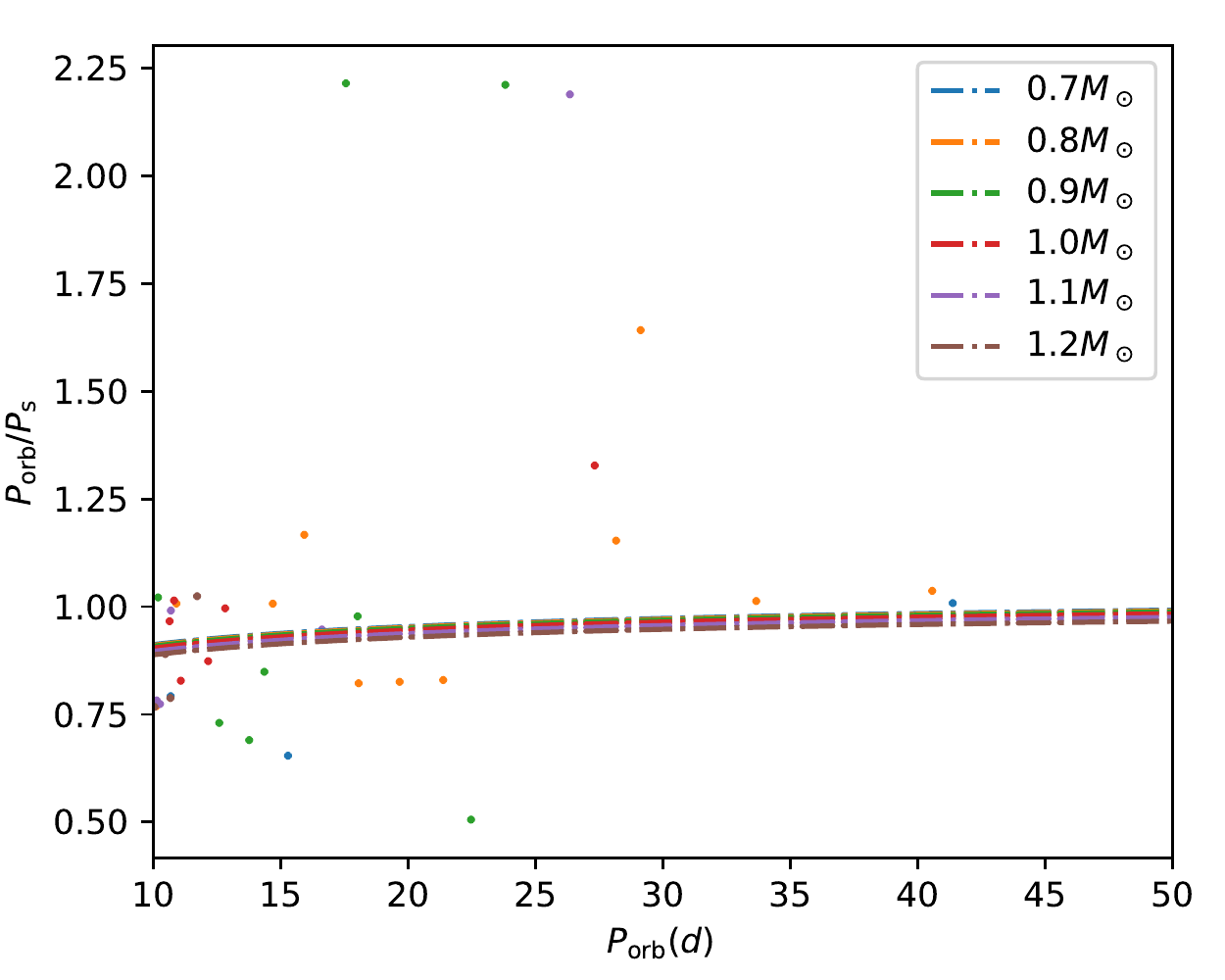}
	\caption{The ratio of orbital period to rotation period is shown as a function of the orbital period and stellar mass, indicated by color. Outliers were identified by the inference procedure as systems with a likelihood at the posterior median parameter values below $10^{-6}$, and are shown as grey circles. These results are from the inference including systems with orbital periods between $10\,{\rm d}$ and $50\,{\rm d}$. The dashed lines indicate the model predictions at the median parameter values for the stellar masses of the same color.}
	\label{fig:1050d}
\end{figure}

\begin{figure}
	\includegraphics[width=0.5\textwidth]{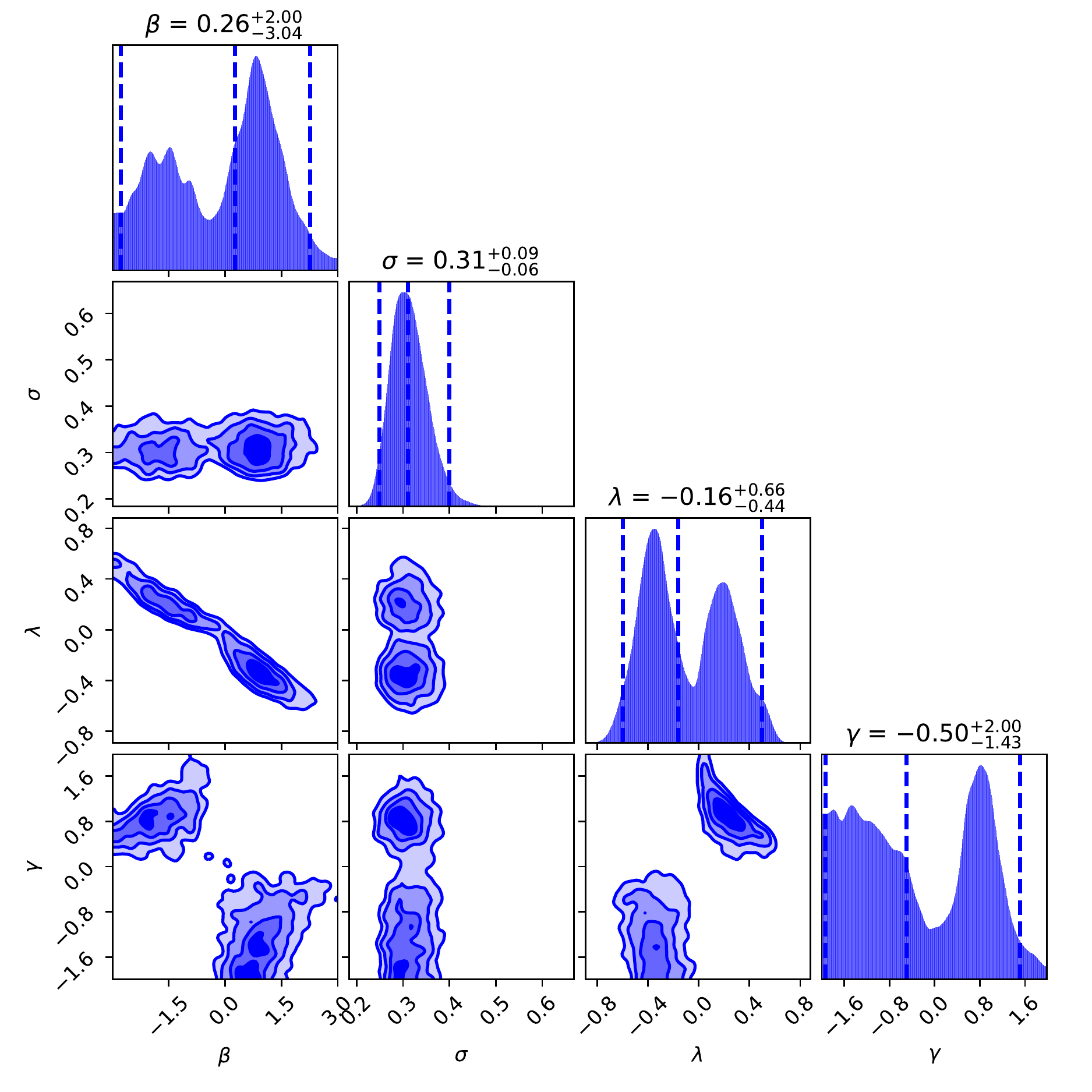}
	\caption{The marginalized posterior distribution over our model parameters is shown for the inference including systems with orbital periods between $10\,{\rm d}$ and $50\,{\rm d}$.}
	\label{fig:1050d_post}
\end{figure}

\begin{figure}
	\includegraphics[width=0.5\textwidth]{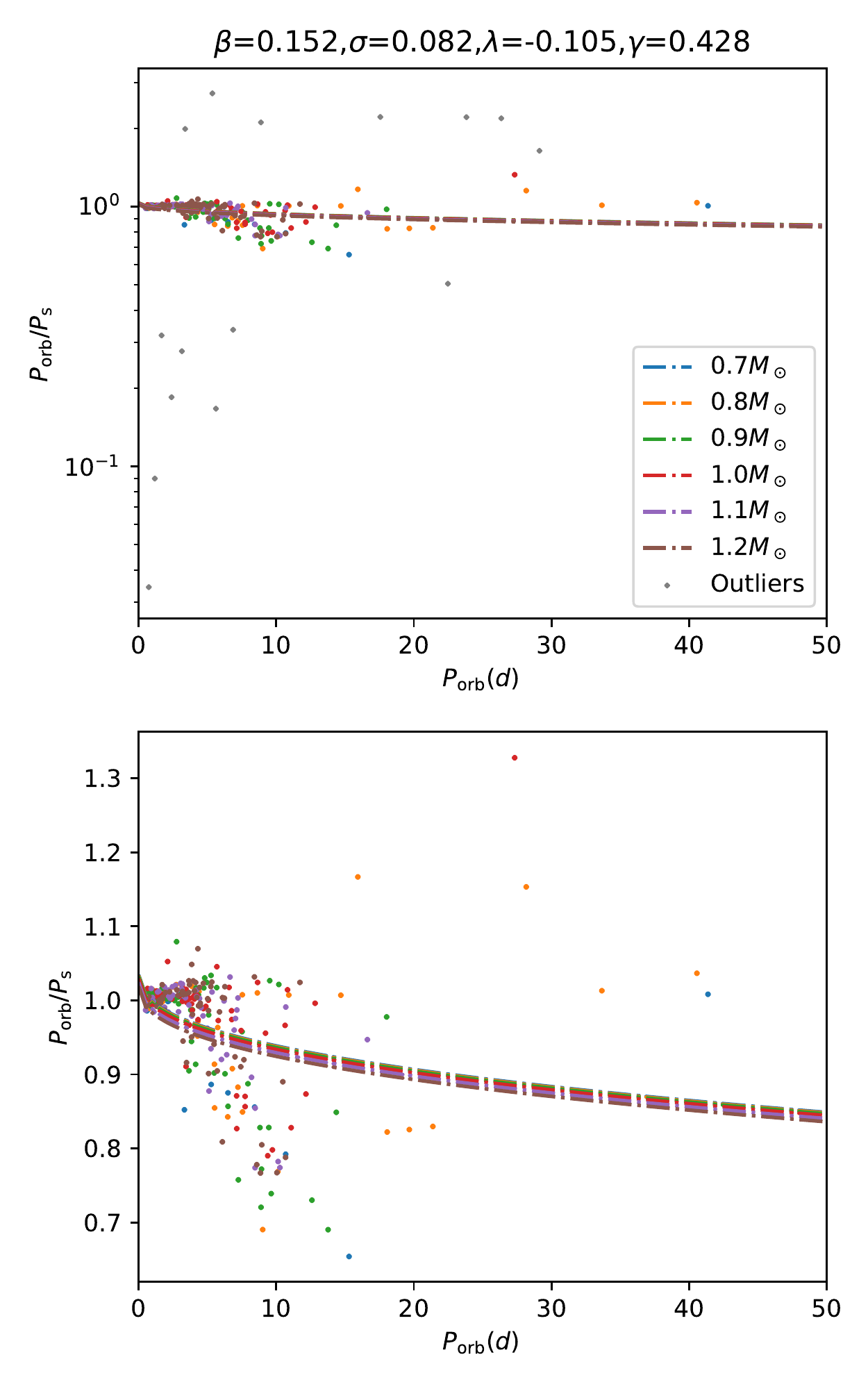}
	\caption{The ratio of orbital period to rotation period is shown as a function of the orbital period and stellar mass, indicated by color. Outliers were identified by the inference procedure as systems with a likelihood at the posterior median parameter values below $10^{-6}$, and are shown as grey circles.  The upper panel shows all objects while the lower excludes outliers. These results are from the inference including systems with orbital periods below $50\,{\rm d}$. The dashed lines indicate the model predictions at the median parameter values for the stellar masses of the same color.}
\end{figure}

\begin{figure}
	\includegraphics[width=0.5\textwidth]{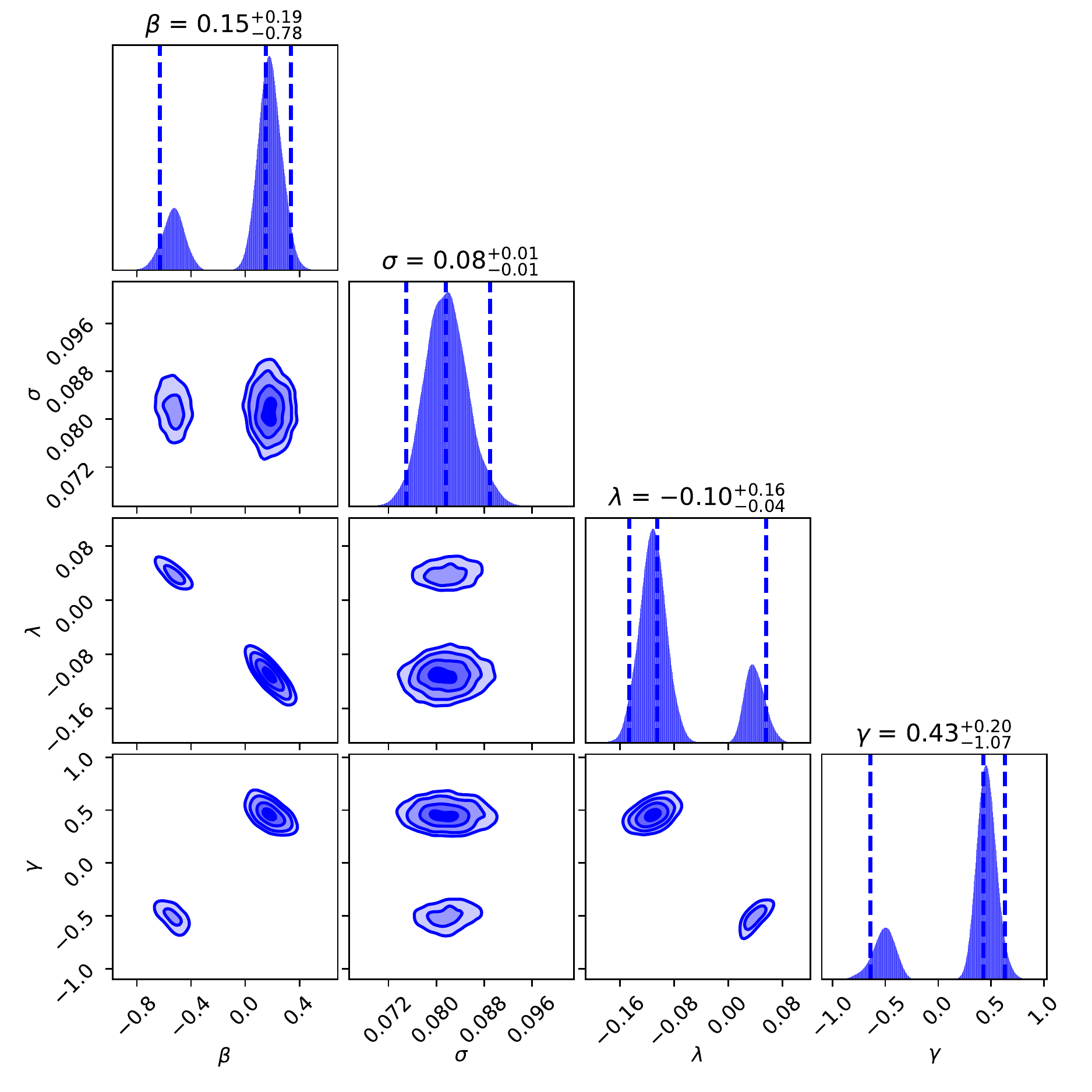}
	\caption{The marginalized posterior distribution over our model parameters is shown for the inference including systems with orbital periods below $50\,{\rm d}$.}
\end{figure}

%%%%%%%%%%%%%%%%%%%%%%%%%%%%%%%%%%%%%%%%%%%%%%%%%%

% Don't change these lines
\bsp	% typesetting comment
\label{lastpage}
\end{document}